\documentclass[10pt,twocolumn,letterpaper]{article}

\usepackage{cvpr}
\usepackage{times}
\usepackage{epsfig}
\usepackage{graphicx}
\usepackage{amsmath}
\usepackage{amssymb}

\usepackage{booktabs}
\usepackage{multirow}
\usepackage{colortbl}
\usepackage[table]{xcolor}
\usepackage{adjustbox}
\usepackage{makecell}

\usepackage[numbers]{natbib}

\usepackage[breaklinks=true,bookmarks=false]{hyperref}

\begin{document}

\newcommand{\XZ}[1]{\textbf{\color{green}[XZ: #1]}} 
\newcommand{\JM}[1]{\textbf{\color{blue}[JM: #1]}} 
\newcommand{\MH}[1]{\textbf{\color{red}[MH: #1]}} 

\newcommand{\unsure}{\textcolor{red}}

\newcommand{\ourmethod}{\textsf{AlbedoEdit}}
\newcommand{\ourdata}{\textsf{AlbedoEdit}}

\newcommand{\timestep}{t}
\newcommand{\cleandata}{\mathbf{x}_\mathrm{0}}
\newcommand{\noisydata}{\mathbf{x}_\mathrm{t}}
\newcommand{\noise}{\epsilon}
\newcommand{\condition}{c}
\newcommand{\velocity}{v}
\newcommand{\encoder}{\mathcal{E}}
\newcommand{\decoder}{\mathcal{D}}

\newcommand{\video}{\mathbf{V}}
\newcommand{\videoalbedo}{\mathbf{A}}

\newcommand{\videozero}{\mathbf{V}^\mathrm{src}}
\newcommand{\videoalbedozero}{\mathbf{A}^\mathrm{src}}
\newcommand{\videoone}{\mathbf{V}^\mathrm{obj}}
\newcommand{\videoalbedoone}{\mathbf{A}^\mathrm{obj}}
\newcommand{\videotwo}{\mathbf{V}^\mathrm{tex}}
\newcommand{\videoalbedotwo}{\mathbf{A}^\mathrm{tex}}

\newcommand{\latent}{\mathbf{z}}

\newcommand{\loss}{\mathcal{L}}

\newcommand{\sv}{\video_\mathrm{s}}
\newcommand{\tv}{\video_\mathrm{t}}
\newcommand{\albf}{\mathbf{a}}
\newcommand{\albfafter}{\albf_\mathrm{1}}
\newcommand{\albfefore}{\albf_\mathrm{0}}

\newcommand{\latentalbedoafter}{\latent^\mathrm{\albfafter}}
\newcommand{\latentalbedobefore}{\latent^\mathrm{\albfefore}}
\newcommand{\latentsv}{\latent^\mathrm{\sv}}
\newcommand{\noisylatenttv}{\latent^\mathrm{\tv}_\timestep}
\newcommand{\cleanlatenttv}{\latent^\mathrm{\tv}}

\newcommand{\normdistribution}{\mathcal{N}}
\newcommand{\denoiser}{\mathbf{f}_\mathrm{\theta}}

\newcommand{\refFig}[1]{Fig.~\ref{fig:#1}}
\newcommand{\refTab}[1]{Tab.~\ref{tab:#1}}
\newcommand{\refSec}[1]{Sec.~\ref{sec:#1}}
\newcommand{\refEq}[1]{Eq.~\ref{eq:#1}}

% SOTA color hacks
\definecolor{myorange}{rgb}{1, 0.85, 0.7}
\definecolor{myred}{rgb}{1, 0.7, 0.7}
\definecolor{myyellow}{rgb}{1, 1, 0.6} % new yellow color

\newcommand{\reducedstrut}{\vrule width 0pt height 1.05\ht\strutbox depth 1.0\dp\strutbox\relax}
\newcommand{\sota}[1]{%
  \begingroup
  \setlength{\fboxsep}{0pt}%  
  \colorbox{myred}{\reducedstrut#1\/}%
  \endgroup
}
\newcommand{\subsota}[1]{%
  \begingroup
  \setlength{\fboxsep}{0pt}%  
  \colorbox{myorange}{\reducedstrut#1\/}%
  \endgroup
}

\newcommand{\thirdsota}[1]{%
  \begingroup
  \setlength{\fboxsep}{0pt}%  
  \colorbox{myyellow}{\reducedstrut#1\/}%
  \endgroup
}

%%%%%%%%%%%%%%%%%%%%%%%%%%%%%%%%%%%%%%%%%%%%%%%%%%%%%%%%%%%%%%%%%%%%%%%%%%%%%%%%
%% Title & Authors
%%%%%%%%%%%%%%%%%%%%%%%%%%%%%%%%%%%%%%%%%%%%%%%%%%%%%%%%%%%%%%%%%%%%%%%%%%%%%%%%

\title{AlbedoEdit: Unified Instance-Level Video Editing with Albedo Guidance}

%% TODO: Replace placeholder author block below with your real authors/affiliations
\author{
  Xilong Zhou$^{1}$ \quad Bao-Huy Nguyen$^{1}$ \quad Zheng Zeng$^{2,3}$  \quad Jacob Munkberg$^{3}$  \quad Jon Hasselgren$^{3}$ \\
  \quad Thomas Leimkühler$^{1}$  \quad Nima Kalantari$^{4}$  \quad Miloš Hašan$^{3}$  \quad Christian Theobalt$^{1}$\\
  \\
  $^{1}$Max Planck Institute for Informatics  \quad $^{2}$University of California Santa Barbara \\
  \quad $^{3}$NVIDIA Research \quad $^{4}$Texas A$\&$M University \\
  % {\tt\small \{author1, author2\}@institution1.edu \quad author3@institution2.edu}
}

\maketitle

%%%%%%%%%%%%%%%%%%%%%%%%%%%%%%%%%%%%%%%%%%%%%%%%%%%%%%%%%%%%%%%%%%%%%%%%%%%%%%%%
%% Teaser figure
%%%%%%%%%%%%%%%%%%%%%%%%%%%%%%%%%%%%%%%%%%%%%%%%%%%%%%%%%%%%%%%%%%%%%%%%%%%%%%%%

\begin{figure*}[t]
  \centering
  \includegraphics[width=\linewidth]{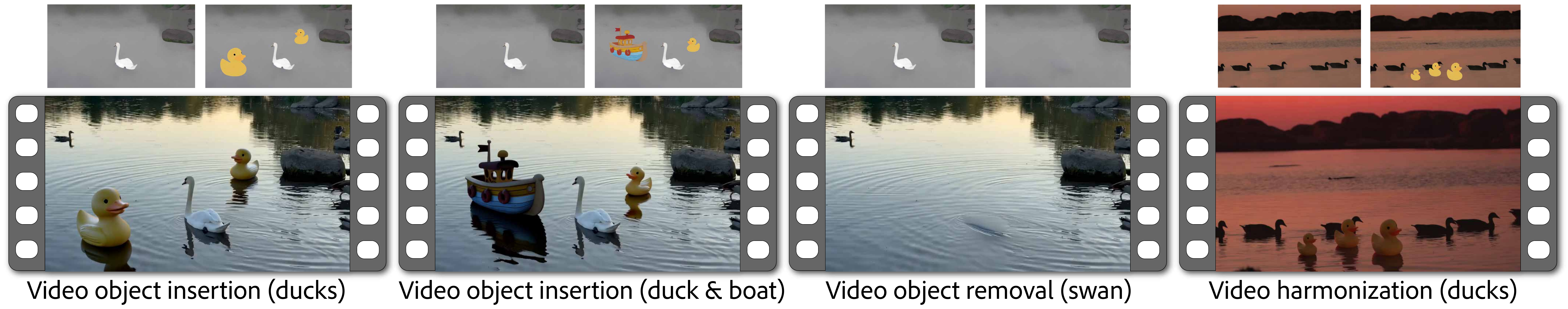}
  \caption{\ourmethod{} is a unified generative video editing framework supporting object insertion, object removal, and texture editing. We leverage the intrinsic albedo map as an effective and user-friendly mechanism for specifying fine-grained appearance edits. We show object insertion, object removal, and an additional insertion example demonstrating lighting harmonization capabilities of the model. The source and target albedo frames are shown above each video.}
  \label{fig:teaser}
\end{figure*}

%%%%%%%%%%%%%%%%%%%%%%%%%%%%%%%%%%%%%%%%%%%%%%%%%%%%%%%%%%%%%%%%%%%%%%%%%%%%%%%%
%% Abstract
%%%%%%%%%%%%%%%%%%%%%%%%%%%%%%%%%%%%%%%%%%%%%%%%%%%%%%%%%%%%%%%%%%%%%%%%%%%%%%%%

\begin{abstract}
Video generative models have achieved remarkable progress in synthesizing photorealistic video sequences. However, enabling broader and more creative downstream applications requires fine-grained instance-level video editing, including object insertion, object removal, and texture editing, which has emerged as a prominent yet challenging problem. Existing approaches either propose unified generative frameworks with only coarse semantic control, or design task-specific frameworks for individual editing tasks, limiting their flexibility and applicability across diverse real-world scenarios. To address these limitations, we propose \ourmethod{}, a unified generative video editing framework that jointly supports object insertion, object removal, and texture editing. Our key insight is that the intrinsic albedo map, which is invariant to lighting and contains no specularity, shadowing and inter-reflection effects, provides an effective and user-friendly mechanism for specifying fine-grained appearance edits. Built upon video foundation models, \ourmethod{} is fine-tuned to translate source RGB videos into edited RGB videos, conditioned on a user-edited first-frame albedo. Trained on a new paired synthetic dataset covering all three editing tasks, \ourmethod{} implicitly learns to harmonize edited contents and simulate complex real-world visual effects triggered by editing operations, including specular highlights, soft shadows, and mirror reflections. \ourmethod{} demonstrates superior performance over state-of-the-art video editing approaches, both qualitatively and quantitatively. Project webpage is \url{https://vcai.mpi-inf.mpg.de/projects/AlbedoEdit}.
\end{abstract}

%%%%%%%%%%%%%%%%%%%%%%%%%%%%%%%%%%%%%%%%%%%%%%%%%%%%%%%%%%%%%%%%%%%%%%%%%%%%%%%%
%% Body
%%%%%%%%%%%%%%%%%%%%%%%%%%%%%%%%%%%%%%%%%%%%%%%%%%%%%%%%%%%%%%%%%%%%%%%%%%%%%%%%

\section{Introduction}

Recent advances in video generative models~\cite{wiedemer2025video, wan2025wan, hong2022cogvideo, kong2024hunyuanvideo, hacohen2026ltx, batifol2025flux} have enabled the synthesis of physically plausible video sequences with high-quality appearance and animation. Despite these advances, extending video generative models to support more flexible and creative downstream applications requires fine-grained object-level video editing. In this paper, we consider three core editing tasks: video object insertion (VOI), video object removal (VOR), and video texture editing (VTE). These tasks aim to manipulate individual objects or regions in a physically accurate manner while preserving the appearance of unedited content (Fig.~\ref{fig:teaser}). They are inherently challenging, as they require a thorough understanding of scene illumination, accurate modeling of physical interactions between elements over time, and faithful simulation of complex visual effects introduced by the edits.

Existing approaches attempt to address these challenges by post-training powerful video generative models. One line of work~\cite{jiang2025vace,wei2025univideo,bian2025videopainter,zi2025cococo} trains unified generative frameworks for multi-modal video editing tasks conditioned on coarse semantic signals. Although these methods can synthesize visually pleasing results, their reliance on coarse semantics prevents them from supporting precise user guidance, such as exact object placement and fine-grained texture manipulation. Another line of work introduces fine-grained object-level control but tackles each editing task independently, such as mask-guided insertion~\cite{gao2026pisco,jin2025insertanywhere}, mask-guided removal~\cite{miao2025rose,fu2026effecterase, motamed2026void}, 3D-aware manipulation~\cite{gu2025diffusion, lee2025generative, yu2025objectmover, wang2025frame}, and intrinsic material editing~\cite{fang2025v, lyu2025intrinsicedit, zeng2024rgb, liang2025diffusion}. Despite achieving higher quality than the unified models, these task-specific frameworks lack flexibility. Moreover, in the VOI task, several of the existing methods tend to preserve the appearance of inserted objects in their original environments, leading to poor harmonization with the source video. 

To address these limitations, we propose \ourmethod{}, a unified video editing framework that jointly supports insertion, removal, and texture editing tasks, while simulating physically plausible complex light transport effects in the edited videos. Our key insight is that the albedo map, as an illumination-invariant intrinsic property, provides a robust conditioning signal for generative video models and offers a simpler avenue for downstream editing operations, compared to direct RGB-space manipulation. Inspired by diffusion-based inverse and forward rendering approaches~\cite{zeng2024rgb,liang2025diffusion,he2025unirelight}, we design \ourmethod{} as an end-to-end generative framework that translates source videos into edited target videos, conditioned on explicitly edited albedo. Specifically, \ourmethod{} is built upon the DiT-based diffusion backbone~\cite{peebles2023scalable}, taking as input the concatenated latents of the source video, the original albedo and the edited albedo, and producing the corresponding edited RGB video through an iterative denoising process.

To train \ourmethod{}, we design and render a large-scale high-fidelity synthetic dataset comprising paired RGB and albedo videos across all three object-level editing tasks, and leverage a Large Vision-Language Model (VLM)~\cite{bai2023qwen} to generate descriptive text prompts for each training sample. During inference, users can simply apply an off-the-shelf inverse rendering model \cite{zeng2024rgb,liang2025diffusion} to extract the first-frame albedo from the source video, apply object-level edits directly to the albedo image, and pass the original and edited albedo frame to \ourmethod{}. Our model implicitly understands the environmental illumination and complex light transport interactions between objects from the source video, plausibly simulating complex visual effects including specular highlights, soft shadows, mirror reflections, and overall harmonization. Despite being trained exclusively on synthetic data, \ourmethod{} generalizes well to diverse real-world scenarios, producing high-fidelity editing results that outperform existing state-of-the-art methods.
Our contributions are summarized as follows:

\begin{itemize}
\item We introduce \ourmethod{}, a unified generative editing framework that jointly supports multiple object-level appearance editing tasks: video object insertion (VOI), video object removal (VOR), and video texture editing (VTE).
\item We demonstrate that the albedo map serves as a powerful tool for object-level video editing, enabling simpler editing operations and illumination-aware simulation of complex real-world visual effects, including specular highlights, soft shadows, and mirror reflections.
\item We build a large-scale, high-quality dataset for three video editing tasks, enabling effective training of an end-to-end editing framework.
\item We demonstrate that the resulting \ourmethod{} model produces high-fidelity video editing results, and outperforms state-of-the-art video editing methods for comparable tasks both qualitatively and quantitatively.
\end{itemize}

We will release our models and dataset to the community.
\section{Related Work}

Built upon successful video generative models~\cite{wan2025wan,batifol2025flux,hong2022cogvideo,yang2024cogvideox,kong2024hunyuanvideo}, controllable video generation and editing have become increasingly critical tasks due to their compatibility with multimodal downstream applications such as content creation, gaming, autonomous driving, and robotics. Controllable video generation covers broad and diverse directions, including personalization~\cite{garibi2025tokenverse,abdal2025dynamic}, camera control~\cite{yu2025trajectorycrafter,he2024cameractrl}, style transfer~\cite{jiang2025vace,gu2025diffusion}, motion control~\cite{lee2025generative,shin2025motionstream}, and object manipulation~\cite{gu2025diffusion}. In this work, we focus on instance-level video editing, and related work in this field can be classified into two categories: unified instance-level editing and task-specific editing.

\subsection{Universal Video Editing}

The first line of work attempts to jointly address multimodal instance-level editing within a unified framework. A popular paradigm adopted by most existing approaches is to extend pretrained diffusion models with auxiliary modality guidance~\cite{zhang2023adding}. Specifically, VACE~\cite{jiang2025vace} proposes to integrate various editing tasks into a Video Condition Unit, incorporated into DiT-based video models to build a universal editing framework. UniVideo~\cite{wei2025univideo} employs a Multimodal Large Language Model for visual understanding alongside a Multimodal DiT for visual generation, forming a dual-stream design. UNIC~\cite{ye2025unic} integrates all input tokens into a single consecutive token sequence. AnyV2V~\cite{ku2024anyv2v} leverages an off-the-shelf image editing model to edit the first frame, and subsequently applies an image-to-video (I2V) model to generate the edited video. Although these unified generative frameworks are capable of synthesizing high-fidelity results, they are typically guided by coarse semantic conditions and lack precise, fine-grained control over instance placement and texture. Moreover, these frameworks struggle to produce photorealistic appearances of edited components, as they do not explicitly disentangle environmental illumination from scene content.

\begin{figure*}[t]
  %\includegraphics[width=\linewidth]{fig/method_c.pdf}
  %\caption{Method overview (both train and inference).}
  \includegraphics[width=\linewidth]{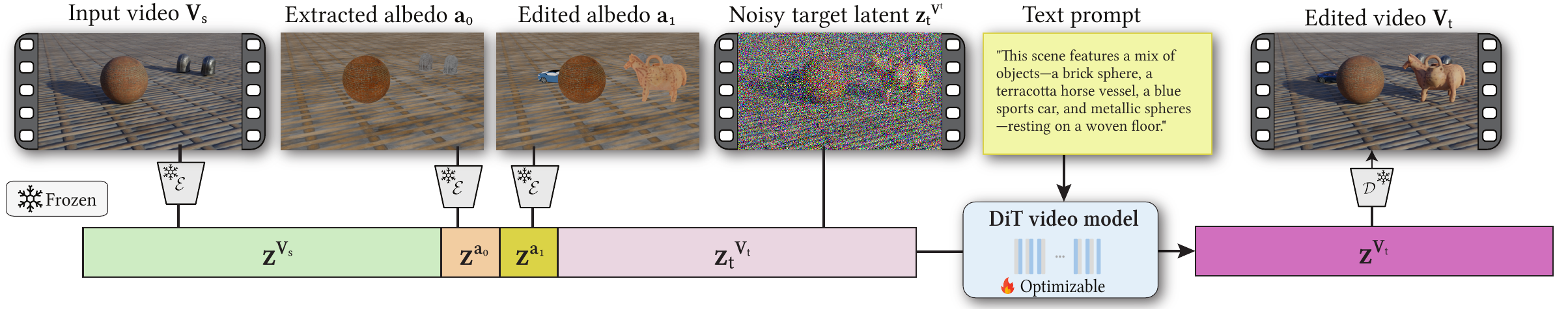}
  \caption{Method overview. Given an input RGB video $\sv$, our video editing framework synthesizes an edited RGB video $\tv$, conditioned on the input video, a text prompt, and two intrinsic albedo images, the extracted albedo $\albfefore$  from the first video frame, and the edited albedo, $\albfafter$. The input video and albedo images are all VAE-encoded to latent space and concatenated along the frame dimension together with the noisy target latent. The text prompt (describing the edited video) is provided through cross-attention, following Wan2.1~\cite{wan2025wan}. The finetuned DiT denoises the latent. Finally, the denoised latent is decoded back to RGB space to produce the edited video.}
  \label{fig:method}
\end{figure*}

\subsection{Task-specific Video Editing}

\textit{Object Insertion} is a prominent yet challenging video editing task. 
Producing plausible edited results requires the framework to not only understand the physical light transport of the source video and image of reference object, but also support user-friendly control, simulating physically accurate visual effects and object harmonization. 
Existing methods tackle this task using different modalities as control signals. Anything in Any Scene~\cite{bai2024anything} inserts and manipulates objects through an object mesh with lighting estimation and shadow generation modules. 
Some existing methods~\cite{zhao2025dreaminsert,saini2024invi, yu2025objectmover} rely on an object mask or instance mask to specify the placement of the inserted object.
To incorporate 3D and 4D priors, InsertAnywhere~\cite{jin2025insertanywhere} leverages a 4D-aware mask generation process, while PISCO~\cite{gao2026pisco} utilizes user-specified keyframe depth maps of the video and reference object image. Although these methods produce convincing edited results, the inserted objects often fail to harmonize correctly with the source video, exhibiting a synthetic appearance.  DiffusionHarmonizer~\cite{zhang2026diffusionharmonizer}, a concurrent work, attempts to address this issue by training diffusion models to refine imperfect composites and improve harmonization.

\textit{Object Removal} aims to erase object instances from a video while also eliminating the secondary effects introduced by the removed object, such as shadow casting, lighting, and reflections. Existing video inpainting methods~\cite{bian2025videopainter,zhou2023propainter, zi2025cococo, li2025diffueraser} can remove certain objects in the source video, but they struggle to eliminate the side effects introduced by the objects. To address this issue, ROSE~\cite{miao2025rose} and EffectErase~\cite{fu2026effecterase} were proposed to automatically learn affected regions outside the mask input and enable side effects removal. These two approaches are effective, but they rely on a mask sequence of the entire video.  A concurrent work, Void~\cite{motamed2026void}, is trained to understand both appearance and physical interaction in the scene, enabling it to delete secondary motion and other side effects introduced by the object.

\textit{Texture Editing} is a long-standing problem that has been extended from the image domain to video. Existing methods~\cite{zeng2024rgb,liang2025diffusion} train separate generative priors for intrinsic decomposition and forward rendering. By editing the extracted intrinsic channels, these approaches can re-render edited scenes under novel illumination conditions. While they support precise material editing, they struggle to fully preserve the input video quality due to ambiguities in illumination estimation. IntrinsicEdit~\cite{lyu2025intrinsicedit} achieves image-level intrinsic editing via DDIM inversion, but this approach is time-consuming and not easily extensible to video. The most recent work V-RGBX~\cite{fang2025v} proposes an end-to-end intrinsic-aware video editing framework supporting fine-grained editing of arbitrary intrinsic channels, but works in two stages (inverse and forward), which struggles to preserve the exact input appearance in unedited areas.

Although task-specific methods produce visually pleasing editing results, their non-unified designs fundamentally limit their flexibility and applicability across diverse real-world scenarios. In contrast, \ourmethod{} is designed as a unified framework that jointly addresses all three instance-level editing tasks, while achieving superior performance over individual task-specific methods.

% \section{Note of related work}

% \subsection{summary}

% Insertion: \textbf{Pisco \cite{gao2026pisco}} + VACE \cite{jiang2025vace} +  \textbf{UniVideo \cite{wei2025univideo}}

% Removal: \textbf{ROSE \cite{gao2026pisco} + EffectErase \cite{fu2026effecterase}} + VideoPainter \cite{bian2025videopainter} + cococ \cite{zi2025cococo}

% Texture Editing: V-RGBX \cite{fang2025v}

% \subsection{code}

% Preserve input video quality + editing from either text prompt or reference image:

% Video Editing Commercial software: NanoBanana/VEO

% Strong baseline Pisco \cite{gao2026pisco}: video editing, insert objects at different frames

% EffectErase \cite{fu2026effecterase}, support insertion and removal

% V2V: VACE \cite{jiang2025vace} +  UniVideo \cite{wei2025univideo}

% ROSE \cite{miao2025rose} only supports removing objects with side effects, not insertion or texture editing

% VideoPainter \cite{bian2025videopainter} and cococ \cite{zi2025cococo}: not conditioned on the reference image; 

% \subsection{ no code}

% InsertAnywhere \cite{jin2025insertanywhere} insert object from reference image to anywhere in the video,

% ObjectMover \cite{yu2025objectmover}: Image space editing, only supports moving objects around, not texture editing or objects insertion

% \subsection{others}

% identity image guided, but with motion trajectory control Frame-in-out \cite{wang2025frame}

% void \cite{motamed2026void} remove both object and interaction, focus more on the motion and interaction
\section{Preliminaries}

Diffusion models~\cite{sohl2015deep,ho2020denoising} have demonstrated state-of-the-art performance in image and video generation tasks~\cite{dhariwal2021diffusion, wan2025wan, kong2024hunyuanvideo, rombach2022high}. The core idea is to approximate a target data distribution through a forward diffusion process and a learned iterative denoising process. In the forward diffusion process, a clean sample $\cleandata \sim q(x)$ is progressively corrupted by adding Gaussian noise $\epsilon \sim \normdistribution(0, \mathbf{I})$, yielding a noisy sample $\noisydata = \alpha_t \cleandata + \sigma_t \epsilon$, where $\alpha_t$ and $\sigma_t$ are time-dependent coefficients determined by noise schedulers. A denoiser $\denoiser$ is then trained to predict the added noise conditioned on the time step $\timestep$, conditional signal $\condition$, and the noisy sample $\noisydata$. The training objective is then formulated as $\left\|\denoiser (x_t, t, c) - \epsilon \right\|_2$. Flow matching \cite{lipman2022flow} offers an alternative formulation in which the forward process is defined as a linear interpolation between the clean data and noise: $\noisydata = (1-\timestep) \cleandata + \timestep \epsilon$. Instead of predicting noise, in the flow matching paradigm $\denoiser$ is trained to predict the target velocity field $\left\|\denoiser (x_t, t, c) - \velocity \right\|_2$, where "velocity" is formulated as $\velocity = \epsilon-\cleandata$. Flow matching is widely used in recent video generation models, and we adopt this paradigm for our video editing framework.

To improve computational efficiency, most video generative models operate in a lower-dimensional latent space, similar to latent diffusion~\cite{rombach2022high}. Specifically, a pretrained Variational Autoencoder (VAE)~\cite{kingma2013auto,van2017neural} $\encoder$ maps an RGB video $\video \in \mathbb{R}^{F \times H \times W \times 3}$, consisting of $F$ frames at spatial reolution $H \times W$, into lower dimensional latent representation $\latent \in \mathbb{R}^{f \times h \times w \times C}$, where $f$, $h$, $w$ denotes downscaled temporal and spatial dimension, and $C$ denotes feature channel. Here, $\denoiser$ operates in latent space, and the denoised latent is decoded into video via the VAE decoder $\decoder$. In addition,  Diffusion Transformer (DiT)~\cite{peebles2023scalable} is one popular video generative framework, where latent space is pachified and flattened into a sequence of tokens and fed into transformer layers. We opt to build \ourmethod{} on the Wan 2.1~\cite{wan2025wan} open-source DiT-based video diffusion model, due to its high-fidelity and temporally consistent video output, as well as ease of use and finetuning.

\section{Method}
\label{sec:inference}

%In this section, we first present our architecture design (\refSec{arch}), and then discuss the key step of dataset construction for the video editing task (\refSec{data}). We present the details of the training and inference process in \refSec{inference}.
In this section, we present our architecture design (\refSec{arch}), describe how we create a synthetic dataset suitable for video editing (\refSec{data}), and discuss the inference (\refSec{inference}) and training processes (\refSec{train}), respectively.

\subsection{Architecture}
\label{sec:arch}

Given an input RGB video $\sv \in \mathbb{R}^{F \times H \times W \times 3}$, the goal of the object-level video editing framework is to synthesize an edited RGB video $\tv \in \mathbb{R}^{F \times H \times W \times 3}$, conditioned on different modalities, such as a text prompt or reference image describing edited objects, or mask indicating the location of the removed or inserted object. In \ourmethod{}, we opt to operate on the intrinsic albedo image $\albf \in \mathbb{R}^{1 \times H \times W \times 3}$ of the first video frame, because of its flexibility for diverse editing tasks (VOI, VOR, and VTE), its invariance to illumination, and its lack of global effects (shadows, reflections). For example, in the VTE task, editing textures in the albedo space allows users to directly replace source textures with target textures without interference from illumination effects. In the VOI task, inserting the delit object albedo facilitates better harmonization and avoids illumination ambiguities introduced by reference images. We empirically find that including albedo before and after editing helps highlight the edited content, producing more robust results than providing only the edited albedo.

The conditions in \ourmethod{} are concatenated as frames to a DiT model. Specifically, the input video $\sv$, along with the first frame albedo before and after editing operations $\albfefore$ and $\albfafter$, are fed into the pretrained VAE encoder $\encoder$ independently, producing the corresponding latents, $\latentalbedobefore$, $\latentalbedoafter$, and $\latentsv$. These latents, together with the time-dependent noisy target latent $\noisylatenttv$, are concatenated along the temporal dimension to form the input to the DiT model. Similar frame-wise concatenation approaches have been used by UniRelight~\cite{he2025unirelight} and MaterialPicker~\cite{ma2025materialpicker}, for relighting and material generation tasks, respectively. We follow the training objective in the flow matching paradigm, and our finetuning objective function is formulated as follows:
\begin{equation}
\label{eq:main_loss}
    \loss = \left\| \denoiser ( \noisylatenttv, \latentsv, \latentalbedobefore,\latentalbedoafter, \timestep, \condition) - \velocity \right\|_2^2,
\end{equation}
where $\condition$ represents descriptive text prompts generated by a Vision-Language Model (VLM)~\cite{bai2023qwen}, and $\velocity$ denotes ground truth velocity parameterized as $\epsilon - \cleanlatenttv$.  Our architecture overview is visualized in~\refFig{method}.

\subsection{Dataset}
\label{sec:data}

\begin{figure}[t]
  \includegraphics[width=\linewidth]{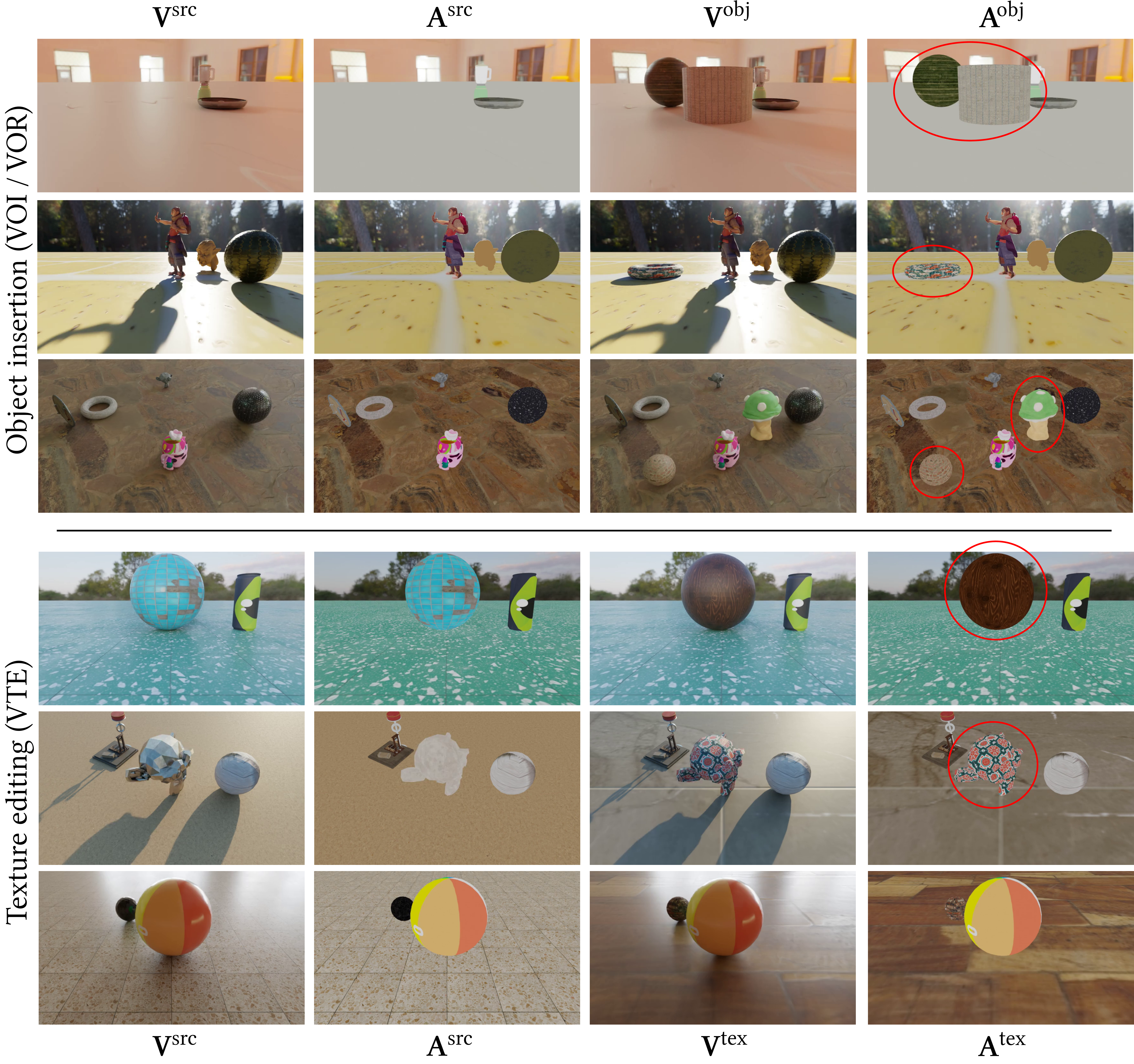}
  %\caption{\XZ{Help} Dataset pipeline and example}
  \vspace{-0.2in}
  \caption{Editing pairs from our training dataset. The top section shows object insertion (and inversely removal) examples and the bottom section illustrates texture editing. Edits are clearly marked in the albedo images except for the last example where the entire ground plane is edited. }
  \label{fig:data}
\end{figure}

\paragraph{Dataset Structure} We construct a synthetic dataset consisting of paired RGB and albedo videos before and after editing operations. Specifically, we organize the data into three groups: ($\videozero$, $\videoalbedozero$) denoting the source RGB and albedo videos without any editing;  ($\videoone$, $\videoalbedoone$) denoting RGB and albedo videos with a set of inserted objects; ($\videotwo$, $\videoalbedotwo$) representing RGB and albedo videos with edited object textures and base colors. Full albedo videos are maintained to allow for clipping and reversal augmentations during training. Training on the source video in combination with the second or the third group enables the system to tackle the corresponding editing tasks. For VOI, the model is trained on the first and second groups ($\videozero$, $\videoalbedozero$, $\videoone$, $\videoalbedoone$), where $\sv = \videozero$, $\tv = \videoone$, $\albfefore = \videoalbedozero[0]$, and $ \albfafter= \videoalbedoone[0]$, following the objective function \refEq{main_loss} described in \refSec{arch}. VOR is essentially the reverse process of VOI, where $\sv$, $\tv$, $\albfefore$, and $\albfafter$ are set as $\videoone$, $\videozero$, $\videoalbedoone[0]$, and $\videoalbedozero[0]$ respectively. Similarly, the model designed for the VTE is trained on the first and third groups ($\videozero$, $\videoalbedozero$, $\videotwo$, $\videoalbedotwo$). In practice, to tackle all these editing tasks jointly, \ourmethod{} is trained on the combinations of all these data within a unified framework. Please refer to \refFig{data} for a selection of editing pairs from our dataset.
%\JM{Perhaps it would be more illustrative to name the three datasets as ($\mathbf{V}^\mathrm{src}$, $\mathbf{A}^\mathrm{src}$), ($\mathbf{V}^\mathrm{obj}$, $\mathbf{A}^\mathrm{obj}$), and ($\mathbf{V}^\mathrm{tex}$, $\mathbf{A}^\mathrm{tex}$).}

\paragraph{Scene Construction}

We follow a similar scene construction strategy as DiffusionRenderer~\cite{liang2025diffusion} and UniRelight~\cite{he2025unirelight} to build diverse synthetic scenes. In each scene, we set up an infinite planar surface with a randomly selected physically-based rendering (PBR) material. On top of the planar surface, we place multiple objects from two sources: $2 \sim 4$ 3D assets curated from Objaverse~\cite{deitke2023objaverse} and $2 \sim 4$ built-in primitives from the rendering engine. To select high-qualty 3D assets from Objaverse, we first manually label 10K objects: assigning a quality score of $1$ to approximately 1K high-quality objects with clean geometry and visually appealing appearance, and a quality score of $0$ to the remaining 9K low-quality objects.  Then we train a lightweight single-layer classifier on the labelled dataset which maps the CLIP embedding~\cite{radford2021learning} of these rendered objects to a scalar quality score. The well-trained classifier, combined with an empirically determined threshold value of $0.8$, is used to filter and select 36K high-quality 3D assets from Objaverse and Objaverse++~\cite{lin2025objaverseplusplus}. 
% \JM{Perhaps cite Objaverse++~\cite{lin2025objaverseplusplus} which proposes something similar.}
In addition to Objaverse objects, we select $2 \sim 4$ geometric primitives (cube, sphere, cylinder, monkey, cone, torus, icosphere) and assign each with a randomly selected PBR material. To better simulate a real-world scenario, our object layout consists of \textit{"hero"} mode, where a primary object is placed at the center of the scene surrounded by smaller secondary objects, and \textit{"non-hero"} mode, in which all objects of similar scale are distributed uniformly across the plane. In both modes, object placement is optimized using bounding box constraints to prevent inter-object intersections. We further introduce diverse smooth camera trajectories and object motions (such as static, rotation, circular translation, etc.) to simulate real-world dynamics. We also apply object scale and rotation augmentations to increase scene diversity.

\begin{figure}[t]
  \includegraphics[width=\linewidth]{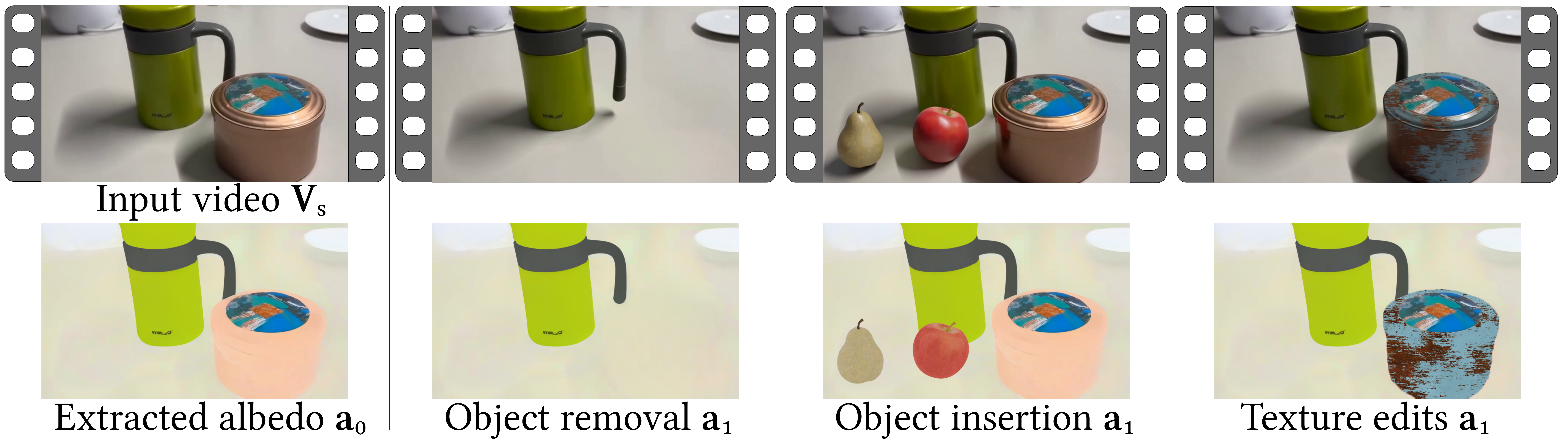}
  \vspace{-0.2in}
  \caption{Our method takes a source video, extracts an albedo image, lets the user apply edits to the albedo image and generates
  new videos following the edits.}
  \label{fig:workflow}
  \vspace{-0.1in}
\end{figure}

\subsection{Inference Pipeline}
\label{sec:inference}

The inference pipeline is visualized in~\refFig{workflow}. Given an input video and an image of the reference object, our goal is to insert the reference object in the video at a specified position, remove existing contents in the video, or edit the texture of specified objects in the video. We first employ the inverse rendering module of an off-the-shelf model, DiffusionRenderer~\cite{liang2025diffusion}, to extract the albedo of the first frame of the video (and optionally the albedo of the inserted object, unless it is a 3D asset whose albedo is simply rendered). Next, we perform task-specific editing operations on the first-frame albedo using a standard image editing tool such as Photoshop~\cite{photoshop}. Specifically, for VOI, the reference object albedo is composited into the scene albedo using alpha blending. For VOR, the target region is removed and filled using built-in image inpainting techniques. Finally, for VTE, the texture of the selected object is replaced with a target texture. The edited albedo image, together with the input video and the  original albedo, is fed into our framework to produce a clean edited video via multiple denoising steps.\looseness=-1

\subsection{Training}
\label{sec:train}

\ourmethod{} is built upon Wan2.1-T2V-14B~\cite{wan2025wan}, and we adopt the default training settings of the model. We fine-tune the full model on 8 H200 GPUs with a batch size of one per GPU and gradient accumulation of 16, resulting in an effective batch size of 128. The model is trained at a resolution of $832 \times 480$ pixels with 33 frames per clip, using the AdamW optimizer~\cite{loshchilov2017decoupled} with a learning rate of $10^{-5}$. We train the model for 30K iterations, equivalent to approximately 2k optimization iterations, which takes approximately eight days. We also conduct experiments on longer video sequences, training a 53-frame model on the same dataset. Note that our architecture uses additional input conditioning (input video and two albedo images) through frame concatenation, so for a video with N latent frames, our 
input latent consists of 2$\times$N+2 latent frames, which roughly doubles the memory and compute requirements compared to the Wan2.1-T2V-14B baseline (which is only conditioned on text). 
% \XZ{report 53 frames experiments briefly and report results in supplementary videos}
%\unsure{Therefore, obtaining a robust 53-frame model requires significantly longer training time, and in this work, all visual and numerical results are reported based on 33-frame models.}

% \MH{We should discuss the frame count trade-offs.}
% \JM{One attempt: Our architecture uses additional input conditioning (input video and two albedo images) through frame concatenation, so for a video with N latent frames, our 
% input latent consists of 2$\times$N+2 latent frames, which roughly doubles the memory and compute requirements compared to the Wan2.1-T2V-14B baseline (which is only conditioned on text)}.

\section{Results}

In this section, we evaluate editing results from our framework, both qualitatively and quantitatively. We first introduce our evaluation protocol in \refSec{eval_proc}, including baseline methods, benchmark datasets, and evaluation metrics. Then we demonstrate comparisons with other baseline methods (\refSec{quant_re} and \refSec{quali_re}) in three editing tasks. Please refer to our supplementary material for editing examples on videos and additional results.

\subsection{Evaluation Protocol}
\label{sec:eval_proc}

\paragraph{Baselines}

\begin{figure}[t]
  \includegraphics[width=\linewidth]{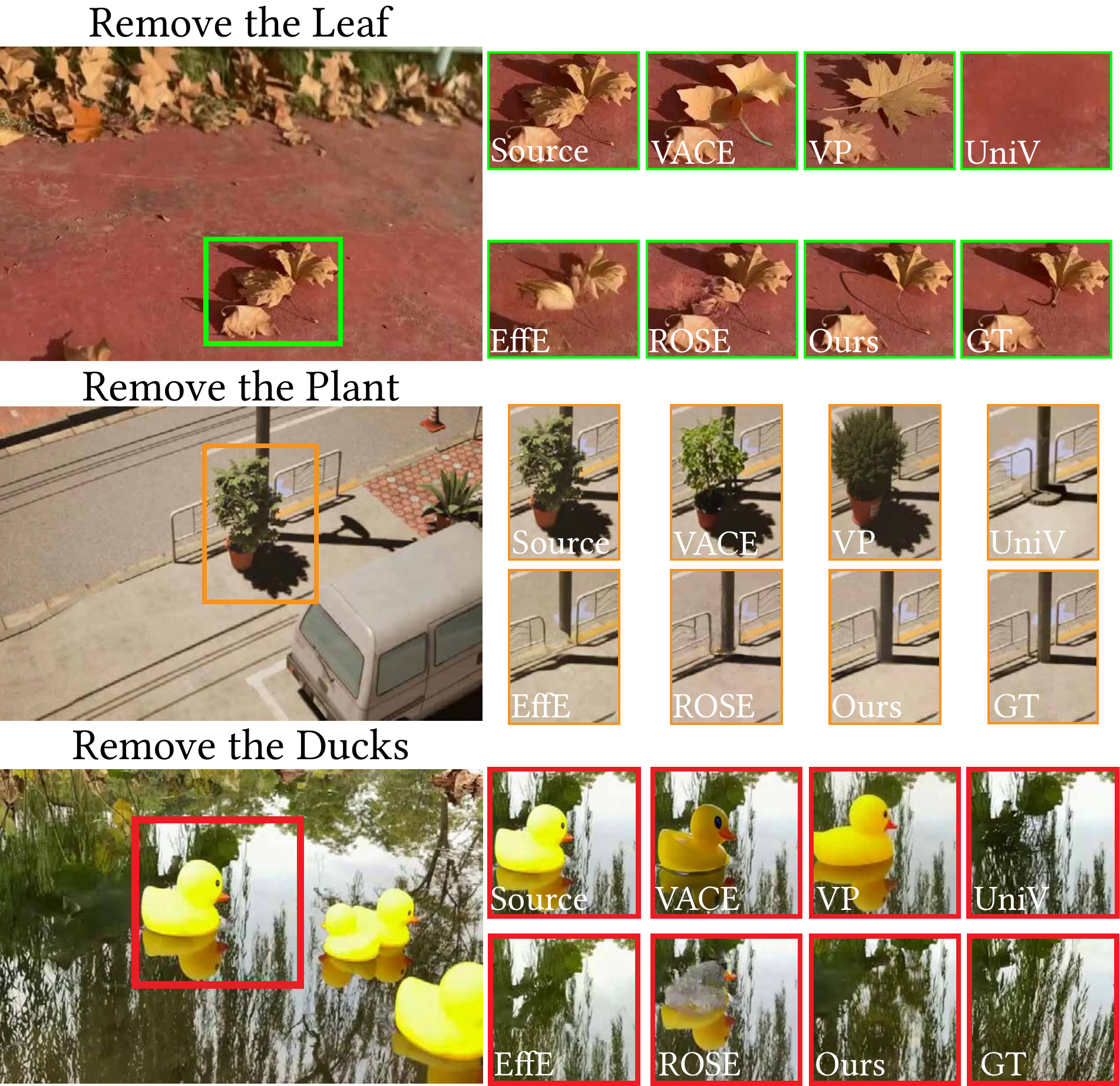}
  \caption{Demonstration of VOR tasks on a benchmark dataset with ground-truth videos. We demonstrate visual results on three object removal examples and compare \ourmethod{} against three universal video editing frameworks (VACE~\cite{jiang2025vace}, VP~\cite{bian2025videopainter}, and UniV~\cite{wei2025univideo}) as well as two methods specifically designed for VOR (EffE~\cite{fu2026effecterase} and ROSE~\cite{miao2025rose}). As shown in the figure, \ourmethod{} achieves the best visual quality in removing objects and their associated side effects under fine-grained control.}
  \label{fig:VOR_gt}
\end{figure}

\begin{figure}[t]
  \includegraphics[width=\linewidth]{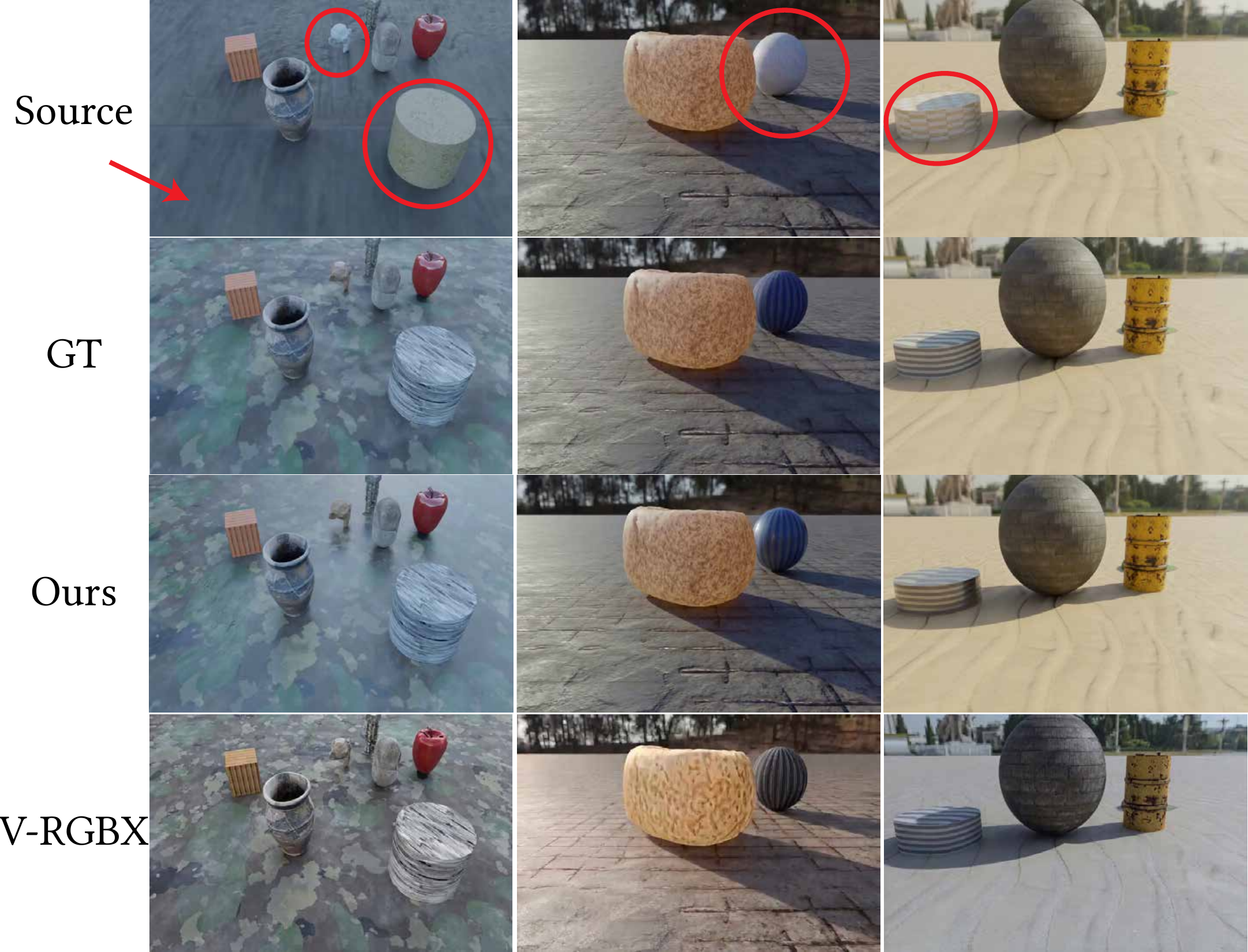}
  \vspace{-0.2in}
  \caption{Demonstration of VTE on a synthetic dataset. Our method is compared against V-RGBX~\cite{fang2025v} conditioned on the edited first frame of albedo in the video. All edited albedo regions are highlighted in the figure, including the ground indicated by the red arrow in the left example and the objects marked with red circles. As demonstrated, compared to V-RGBX, our method preserves the visual quality of the source video while restricting changes only to the intended edited regions, yielding coherent and physically plausible results. V-RGBX, as a two-stage inverse-forward render method, leads to slight changes in all image areas.}
  \label{fig:VTE_gt}
  \vspace{-0.2in}
\end{figure}

% VOI: Pisco + VideoPainter + VACE + UniVideo

% VOR: EffectEraser + ROSE + VACE + VideoPainter

% VACE + UniVideo: 33 frames

% VideoPainter: 49 frames

% EffectEraser run on 81 frames: 832x480

% ROSE run on 49 frames: 720x480

% Pisco run on 49 frames: 720x480

% V-RGBX: 53 frames

We compare \ourmethod{} against previous unified frameworks designed for multiple video editing tasks, as well as task-specific methods for VOI, VOR, and VTE. For unified video editing frameworks, we compare against VACE~\cite{jiang2025vace}, UniVideo (UniV)~\cite{wei2025univideo}, and VideoPainter (VP)~\cite{bian2025videopainter}. For task-specific baselines, we compare against ROSE~\cite{miao2025rose} and EffectErase (EffE)~\cite{fu2026effecterase} for VOR, PISCO~\cite{gao2026pisco} for VOI, and V-RGBX~\cite{fang2025v} for VTE. All methods are evaluated using the official code released by the authors. To ensure a fair comparison, all experiments are conducted at a similar resolution to our method. For baseline methods that require a text prompt as input, we utilize the same descriptive text prompt as our method. We test PISCO conditioned on the first frame only, consistent with our method.

\begin{figure*}[t]
  \includegraphics[width=\linewidth]{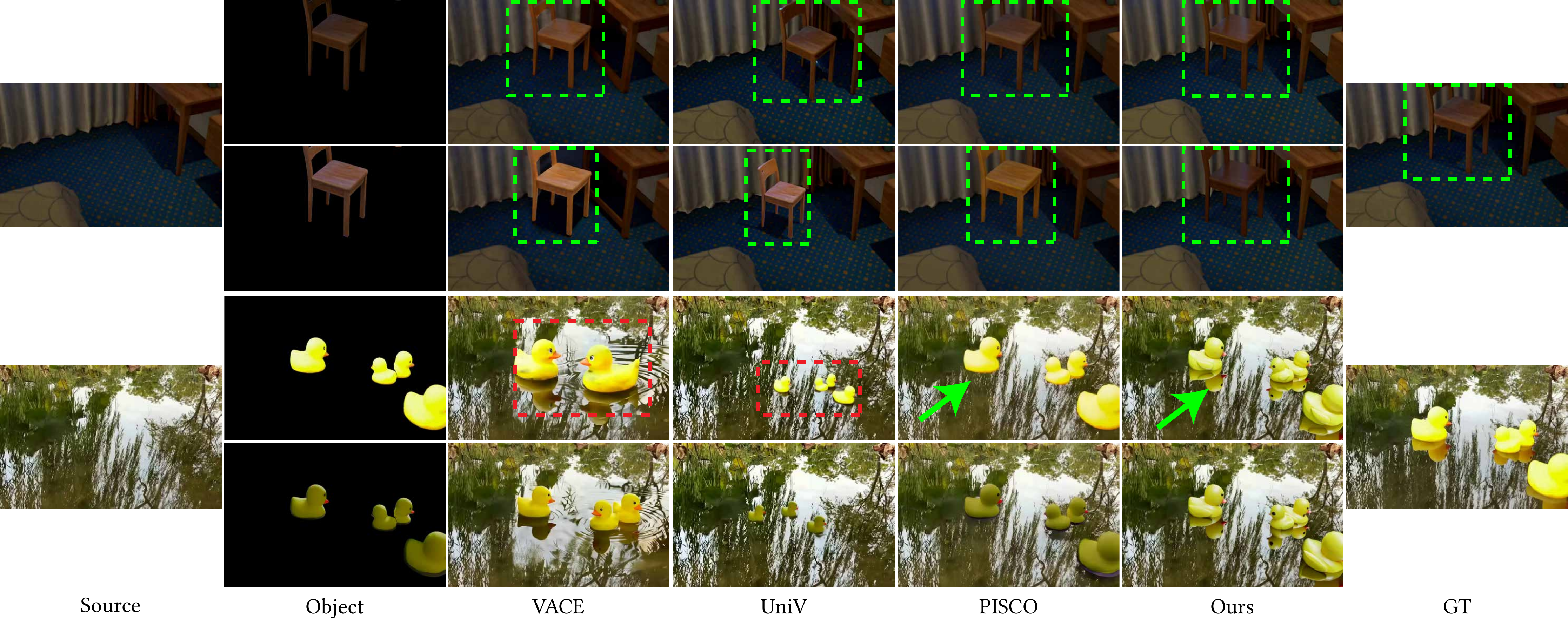}
  \vspace{-0.2in}
  \caption{Demonstration of VOI tasks on the benchmark dataset with GT videos. Our method is compared against three universal video editing frameworks (VACE~\cite{jiang2025vace}, VP~\cite{bian2025videopainter}, UniV~\cite{wei2025univideo}) and one VOI task (PISCO~\cite{gao2026pisco}). We show the insertion of the original objects as well as one relit object (VOI-Relit) into the source video. Existing unified frameworks lack precise control over the inserted objects, such as position and scale (indicated by the dashed square), whereas both PISCO and \ourmethod{} can produce fine-grained control of inserted objects. However, PISCO struggles to achieve physically plausible harmonization for the relit objects and misses the mirror reflection on the lake (indicated by the green arrow). Overall, \ourmethod{} produces high-quality, physically plausible edits while maintaining precise control over object insertion.}
  \label{fig:VOI_gt}
\end{figure*}

\paragraph{Evaluation Dataset}

To conduct an extensive evaluation, we demonstrate both quantitative and qualitative results on two groups of datasets: 1) datasets with ground truth (GT) or pseudo-GT, and 2) datasets without GT. For datasets with GT, we curate 9 scenes from the ROSE benchmark~\cite{miao2025rose} and 7 scenes from VOR-Eval~\cite{fu2026effecterase}, both of which contain synthetic and real video pairs captured before and after object removal. We exclude videos exhibiting blurriness, repetitive content, defocus, strong motion caused by edited objects, light source editing tasks, or objects with ambiguous identity or overly complex structure. 

We evaluate both the VOR and VOI tasks on the resulting dataset, given that these two processes are inverses of one another. However, in practice, inserted objects should not share the same lighting environment as the background video. To better evaluate robustness and simulate real-world applications, we perform a "relit VOI", where the same object is inserted after being relit using DiffusionRenderer~\cite{liang2025diffusion} under different illumination. For VTE evaluation, we render 10 synthetic scenes following our dataset reconstruction pipeline with albedo edits. In addition to these ground-truth datasets, we qualitatively evaluate on in-the-wild captured data. We curate 24 samples from VOR-Wild~\cite{fu2026effecterase} and several videos captured by ourselves. We evaluate VOR on the VOR-Wild data by removing objects present in the original videos, and further assess VOI and VTE by inserting external objects, which do not exist in the original video, and performing texture editing on these in-the-wild videos.\looseness=-1

\paragraph{Evaluation Metrics}

We evaluate edited video quality using standard image metrics: Peak Signal-to-Noise Ratio (PSNR), Structural Similarity Index Measure (SSIM)~\cite{wang2004image}, and Learned Perceptual Image Patch Similarity (LPIPS)~\cite{zhang2018unreasonable}, computed on benchmark datasets with ground truth. For VOR and VTE, metrics are computed over full edited sequences; for VOI, evaluation is performed solely on the first frame, since our method takes only the first-frame albedo as input and provides no motion guidance. For evaluation across different methods, we crop and resize the GT videos to match the training resolution of each method, following its corresponding video processing pipeline. In addition to GT-based metrics, we employ VBench~\cite{huang2023vbench} to assess comprehensive video quality for VOR and VOI evaluation on the ROSE and VOR-Eval datasets.

\begin{figure*}[tbp]
  \includegraphics[width=\linewidth]{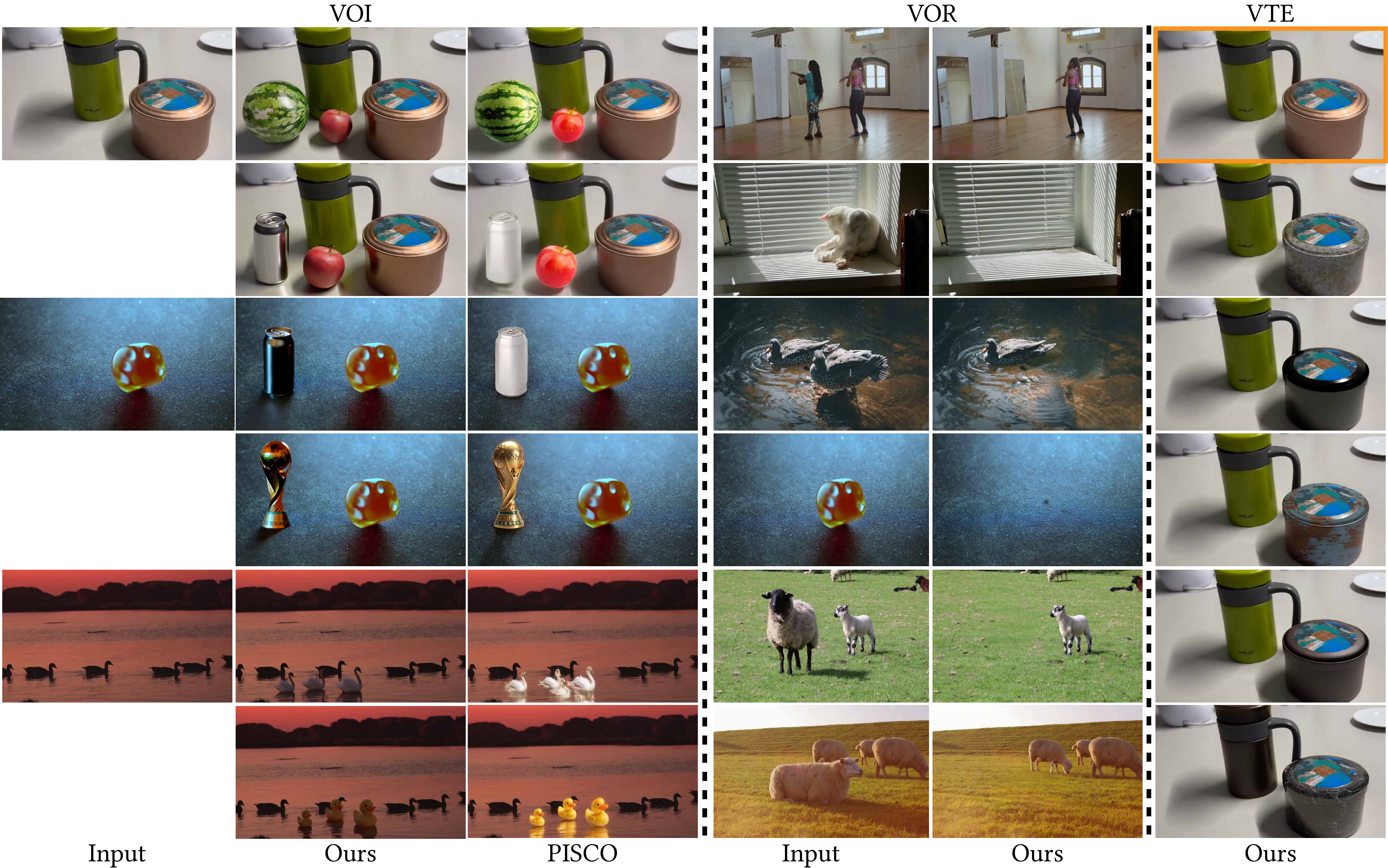}
  \vspace{-0.2in}
  \caption{Demonstration of free-form edits on in-the-wild videos. From left to right, we demonstrate VOI, VOR, and VTE results, respectively. Input source videos in the VTE task are highlighted with an \textcolor{orange}{orange} border. For VOI (left), we insert external objects and compare our method qualitatively against PISCO~\cite{gao2026pisco}. \ourmethod{} achieves physically plausible harmonization with the scene, and correctly guesses which objects should have metallic appearance, whereas PISCO preserves the original appearance of the inserted objects and struggles to relight the object under the source video environment. For VOR (middle), \ourmethod{} effectively removes objects along with their complex side effects, even for objects with strong motion (e.g., rotating dice and birds interacting with water). For VTE (right), by replacing the texture of real-world objects with synthetic albedo maps, \ourmethod{} resynthesizes objects with photorealistic appearance and plausibly guesses material parameters unspecified in the albedo, such as a tea box made of glossy marble or leather.}
  \label{fig:results_wild}
\end{figure*}

\subsection{Quantitative Results}
\label{sec:quant_re}

We demonstrate numerical results on the evaluation datasets with GT in \refTab{num} and visual results for three different tasks in \refFig{VOR_gt}, \refFig{VOI_gt}, and \refFig{VTE_gt}, respectively. As shown in \refTab{num}, \ourmethod{} ranks among the top two across different editing tasks, achieving results quantitatively comparable to methods tailored for a specific task and numerically superior to most universal editing methods. Note that although VACE achieves a higher VBench score than ours on the VOR task, it typically only generates plausible videos under coarse control and lacks fine-grained control. For example, in \refFig{VOR_gt}, VACE does not correctly remove the objects (leaf, plant, and ducks), which can still yield a high VBench score yet a low score on GT-based metrics. We highlight that in the first VOR example, \ourmethod{} is the only method that correctly removes the small leaf piece, owing to the strong constraints of the albedo map.

For VOI, PISCO performs best on GT-based metrics since it is tailored for the insertion task, and the standard VOI task involves inserting objects that originally existed in the video. The appearance of the inserted object in the standard VOI task is already harmonized with the background, and PISCO only focuses on capturing correct side effects such as shadows and reflections. In comparison, \ourmethod{} learns to understand the background environment, relight the inserted objects, and reproduce the side effects they introduce. In addition to standard VOI, we also evaluate \textit{VOI-Relit}, as discussed in \refSec{eval_proc}, where the framework is conditioned on the reference image of the same relit object under novel illuminations. As shown in the table, in the \textit{VOI-Relit} experiment, we achieve better PSNR and VBench scores and comparable SSIM to PISCO, while obtaining worse LPIPS. We conclude that PISCO is good at capturing high-frequency details of the object, to which LPIPS is sensitive, but fails to perform correct harmonization and color tone adjustment, resulting in worse PSNR than ours. As demonstrated in \refFig{VOI_gt}, universal methods can insert objects into videos, but they typically lack control over object position and size. While PISCO achieves visually plausible results for the original VOI task, it produces incorrect harmonization when inserting relit objects (the chairs and ducks). In comparison, \ourmethod{} not only achieves precise position control, but also supports better harmonization and relighting than other approaches.\looseness=-1

For the VTE task, we compare against V-RGBX~\cite{fang2025v} using an edited first-frame albedo map. As demonstrated both qualitatively and quantitatively in \refTab{num} and \refFig{VTE_gt}, although V-RGBX can achieve decent albedo editing, it struggles to preserve the appearance of the source video. In comparison, \ourmethod{} preserves the appearance and illumination of the source video while also maintaining correct albedo editing introduced by the albedo maps.

\subsection{Qualitative Results}
\label{sec:quali_re}

In addition to quantitative evaluations on datasets with GT, we perform qualitative evaluations on an in-the-wild dataset and conduct free-form edits (without ground truth) for the three tasks, as demonstrated in \refFig{results_wild}. Specifically, we show in-the-wild VOI, inserting external objects that do not exist in the source video, VOR, removing challenging \emph{moving} objects in real videos, and VTE, replacing real object textures with different synthetic albedo maps. We also qualitatively compare \ourmethod{} against PISCO for the in-the-wild VOI task in \refFig{results_wild}. As shown, PISCO struggles with external object insertion, producing poor harmonization with synthetic appearance, such as the white swan and yellow ducks inserted into a reddish sunset scene. In comparison, \ourmethod{} demonstrates physically plausible relighting and harmonization and can even capture physically plausible specular reflection interactions between inserted objects and original objects in the source video, such as the reflection of fruits on the surface of the tea cup in the top example and the reflection of the die on the surface of the inserted can or trophy in the middle example of the VOI task. In addition, our VOR and VTE results are also physically convincing, e.g., replacing the texture of a tea cup with glossy marble in the bottom VTE example. We strongly recommend that readers consult the supplementary videos for a complete presentation of our approach. In summary, \ourmethod{} demonstrates strong and robust editing results both qualitatively and quantitatively within a unified editing framework.

\begin{table}[t]
    \renewcommand{\arraystretch}{0.80}
    \centering
    \caption{Quantitative comparison of our method against baselines across three tasks (VOR, VOI, and VTE). We utilize three standard GT-based metrics (PSNR, LPIPS, and SSIM), and the weighted average score (Avg) of VBench for video quality evaluation without GT. In addition to VOI, we also include numerical evaluation on the \textit{VOI-Relit} task (where the object is relit before insertion).}
    \label{tab:num}
\adjustbox{max width=\textwidth}{
{\setlength{\tabcolsep}{3pt}  % default is 6pt
\begin{tabular}{@{}ll @{\hspace{6pt}} ccc @{\hspace{6pt}} c@{}}
    \toprule
    \multirow{2}{*}{Task} & \multirow{2}{*}{Method}
        & \multicolumn{3}{c}{GT-based Metrics}
        & \makecell{VBench} \\
    \cmidrule(lr){3-5} \cmidrule(lr){6-6}
        & & PSNR$\uparrow$ & SSIM$\uparrow$ & LPIPS$\downarrow$ & Avg$\uparrow$ \\
    \midrule
    % ── VOR ──────────────────────────────────────────────────────────────
    \multirow{6}{*}{\textit{VOR}}
        & VACE          & 22.37 & .866 & \thirdsota{.246} & \sota{.662} \\[-1pt]
        & UniV          & 21.00 & .685 & .368 & .645 \\[-1pt]
        & VP            & 21.05 & .839 & .311 & \thirdsota{.655} \\[-1pt]
        & EffE          & \subsota{27.84} & \thirdsota{.899} & .256 & .652 \\[-1pt]
        & ROSE          & \sota{28.17} & \sota{.919} & \subsota{.244} & .648 \\[-1pt]
        & \textbf{Ours} & \thirdsota{27.64} & \subsota{.914} & \sota{.220} & \subsota{.656} \\
    \midrule
    % ── VOI ──────────────────────────────────────────────────────────────
    \multirow{5}{*}{\textit{VOI}}
        & VACE          & \thirdsota{23.15} & \thirdsota{.858} & \thirdsota{.248} & \thirdsota{.658} \\[-1pt]
        & UniV          & 17.99 & .561 & .430 & \subsota{.660} \\[-1pt]
        & VP            & 21.17 & .848 & .313 & .656 \\[-1pt]
        & PISCO         & \sota{27.10} & \sota{.912} & \sota{.205} & .655 \\[-1pt]
        & \textbf{Ours} & \subsota{25.22} & \subsota{.895} & \subsota{.233} & \sota{.661} \\
    \midrule
    % ── VOI-Relit ────────────────────────────────────────────────────────
    \multirow{4}{*}{\textit{VOI-Relit}}
        & VACE          & 22.44 & .853 & .250 & \subsota{.664} \\[-1pt]
        & UniV          & 17.58 & .557 & .434 & .662 \\[-1pt]
        & PISCO         & \subsota{24.38} & \sota{.891} & \sota{.219} & .660 \\[-1pt]
        & \textbf{Ours} & \sota{24.79} & \sota{.891} & \subsota{.236} & \sota{.672} \\
    \midrule
    % ── VTE ──────────────────────────────────────────────────────────────
    \multirow{2}{*}{\textit{VTE}}
        & V-RGBX        & 19.46 & .807 & .378 & -- \\[-1pt]
        & \textbf{Ours} & \sota{28.04} & \sota{.903} & \sota{.217} & -- \\
    \bottomrule
\end{tabular}
}}
\end{table}

% \subsection{Notes}

% \paragraph{****Insetion Dataset}

% 1. ROSE-Evaluation + VOR-evaluation: 

% precise control: ours (1st frame) vs Pisco (1st frame)--> lpips, psnr, fvd with GT + VBench w/o GT
% 1) insert original natural RGB instance
% 2) insert relit natural RGB instance

% not precise control:ours (1st frame) vs UniVideo + VACE (text/img guided) -->  VBench w/o GT
% 1) insert original natural RGB instance
% 2) insert relit/albedo RGB instance

% 2. Several real videos?

% PISCO-Bench and VOIBench not released, so we need to reconstruct by ourselves, from several real scene, remove obj using ROSE or EffectEraser, then reinject.

% \paragraph{****Removal Dataset}

% ROSE-Evaluation + VOR-evaluation with GT: ours (1st mask) vs other (video mask) --> lpips, psnr, fvd, etc

% \paragraph{****Texture Editing Dataset}

% Use our rendered 10 synthetic scenes as inference + other potential dataset?

% compared with V-RGBX with albedo edited.

% \begin{figure}[t]
%   \includegraphics[width=\linewidth]{example-image}
%   \caption{Limitation}
%   \label{fig:limitation}
% \end{figure}
\subsection{Additional Experiments}

% In this section, we first discuss robustness against illumination changes of inserted objects in VOI, followed by an ablation study on the impact of the size of our video base model. We conclude with limitations and future work.

\paragraph{Effect of emission}

\begin{figure}[t]
  \includegraphics[width=\linewidth]{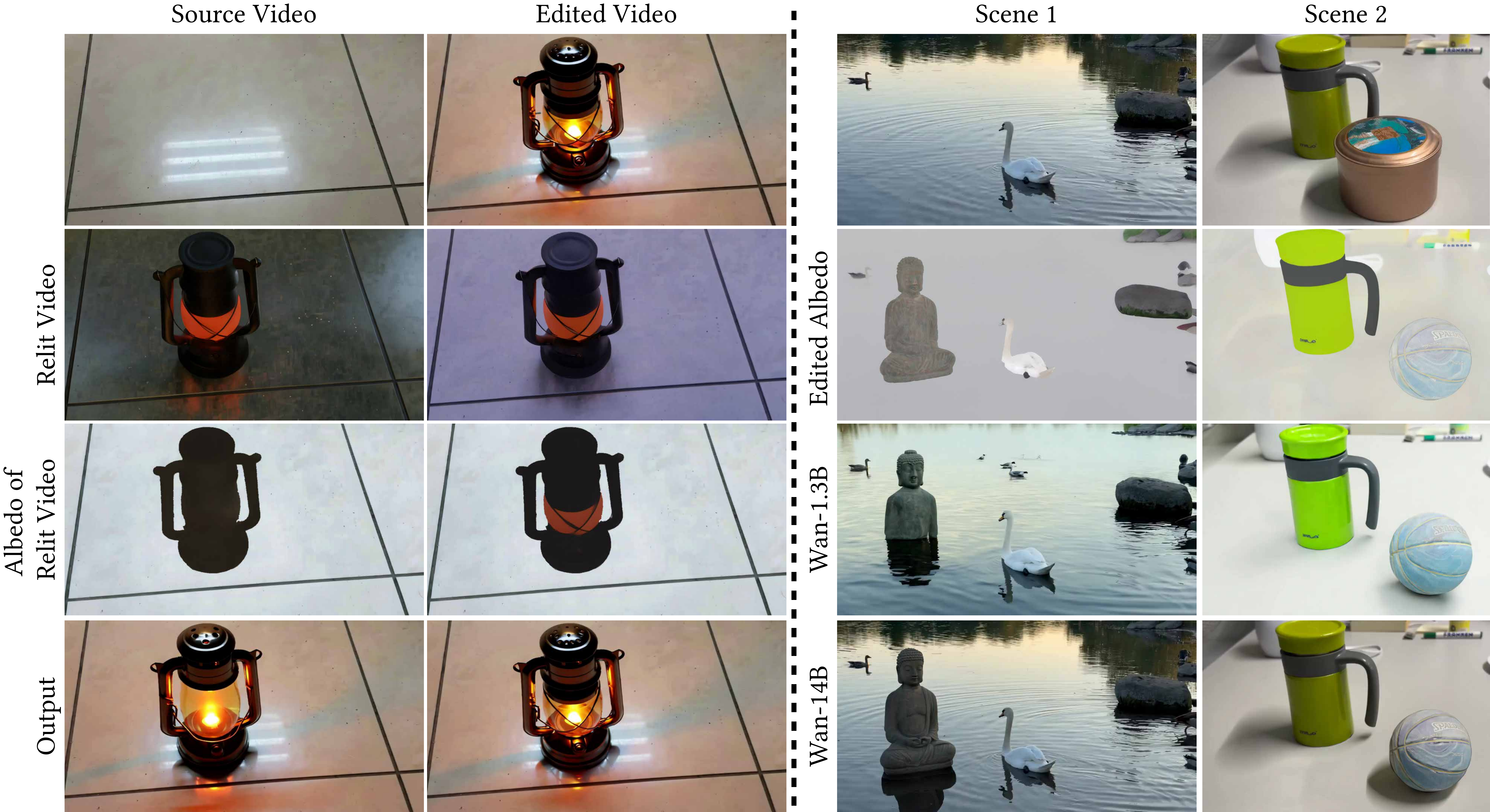}
  \vspace{-0.2in}
  \caption{Left: We demonstrate the effect of inserting objects that should plausibly be emissive; the model correctly infers this. Right: using a smaller base model (Wan 1.3B) on the insertion task leads to noticeably lower quality.}
  \label{fig:ablation}
\end{figure}

We study the effect of illumination on inserted objects in the \textit{VOI-Relit} task. As demonstrated in the left part of \refFig{ablation}, we relight the edited video using DiffusionRender~\cite{liang2025diffusion} to obtain relit videos, from which albedo maps are extracted and utilized for another round of VOI. In the example shown in \refFig{ablation}, even though the extracted albedo of the lantern differs between the two examples, the final outputs match the original edited video and the inserted relit objects remain physically plausible. Additional VOI-Relit results are presented in \refFig{VOI_gt}, where relit objects (e.g., chairs and ducks) are naturally harmonized into the source video. These experiments demonstrate the effectiveness of \ourmethod{} in tackling in-the-wild external object insertion.

\paragraph{Effect of base model size}

We evaluate different video foundation models on our editing tasks. Specifically, we fine-tune the smaller Wan-1.3B-T2V model on the same dataset; note that our main results
use the larger fine-tuned Wan-14B-T2V model. As shown in the right part of \refFig{ablation}, the model fine-tuned from Wan-1.3B struggles to preserve the appearance of the source video, highlighting the benefits of larger models for such tasks. While larger models are more powerful, they also require more memory.

\subsection{Limitations and Future Work}

\begin{figure}[t]
  \includegraphics[width=\linewidth]{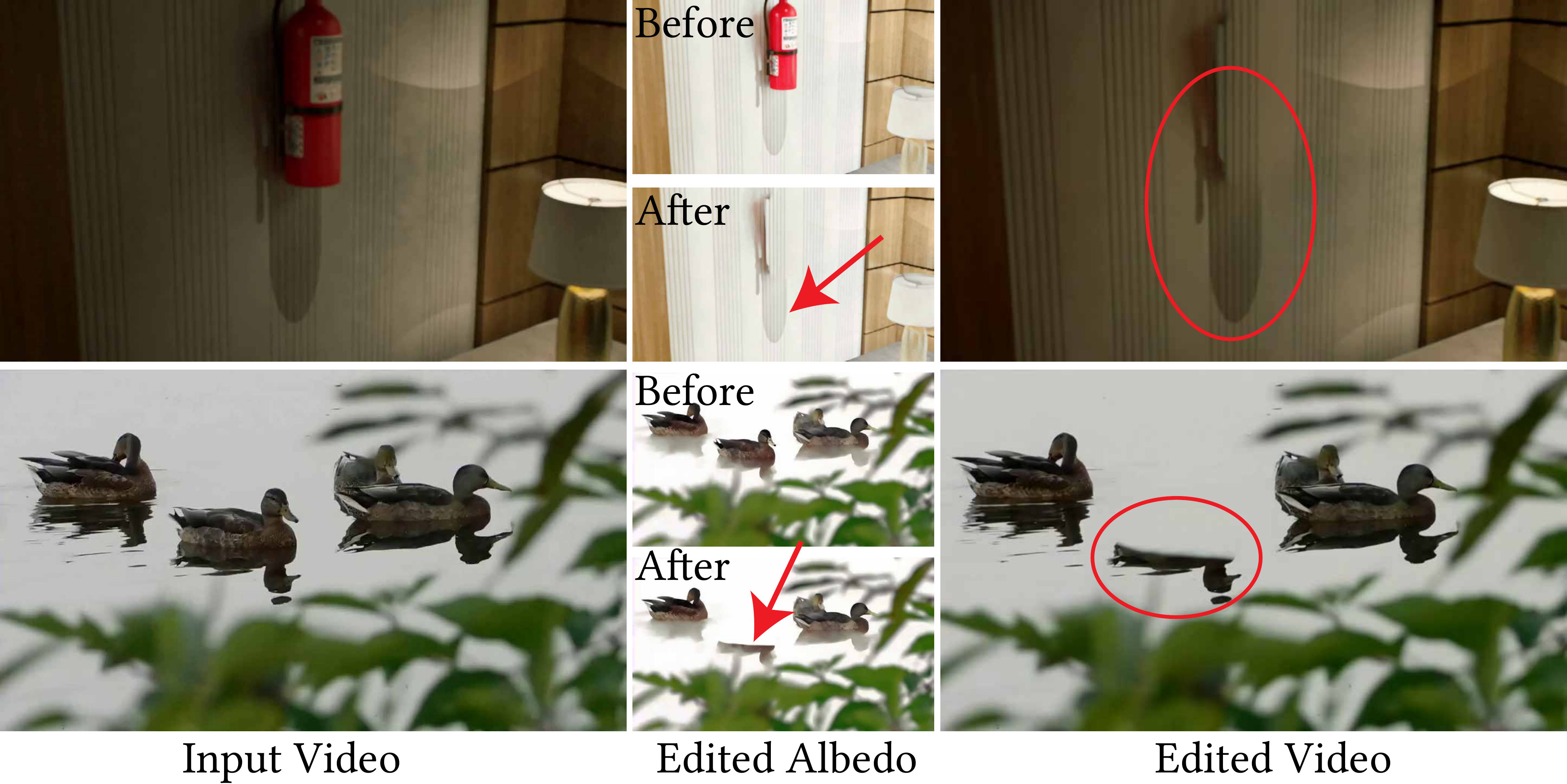}
  \vspace{-0.2in}
  \caption{Limitation of our method. As shown in the figure, when side effects of edited objects are mistakenly counted as texture (highlighted by the red arrow) in the albedo maps, our method fails to remove the corresponding side effects in the edited videos.}
  \label{fig:limitation}
\end{figure}

Our albedo guidance is currently applied to the first frame only. This is often beneficial, as editing the albedo of one frame is much easier for a user than the concurrent modification of multiple albedo frames with temporal coherence. However, it also prevents precise motion control; adding such control is an important remaining challenge. In addition, the plausibility of the synthesized videos relies on the quality of the extracted albedo maps. In cases where reflections or shadows of the object are mistakenly interpreted as albedo, \ourmethod{} may fail to remove the corresponding side effects, as shown in \refFig{limitation}.

Our current model supports short videos only, and our results use 50 diffusion steps, with no distillation. Autoregressive long-video approaches with distillation have been heavily explored recently and could be combined with our approach.

\section{Conclusion}

In this paper, we introduced \ourmethod{}, a unified framework for several video editing tasks: inserting new objects (VOI), removing existing ones (VOR), and changing textures (VTE). Basing our guidance on albedo maps, which represent an object's intrinsic appearance without the distraction of shadows and global illumination, allows the model to preserve the lighting of the original scene, ensuring that new edits include natural details like accurate shadows, highlights, and reflections that remain difficult for other methods.

Our results show that \ourmethod{} produces comparable or higher quality and more consistent videos than previous specialized tools, which typically focus on one of the three tasks we support. Due to our careful construction of synthetic training data, the model generalizes well to real-world footage. We showed that the albedo map, an illumination-invariant intrinsic property, provides a robust conditioning signal to guide the editing process. We believe this approach can inspire future tools to make generative video editing accessible, controllable and photorealistic.

%%%%%%%%%%%%%%%%%%%%%%%%%%%%%%%%%%%%%%%%%%%%%%%%%%%%%%%%%%%%%%%%%%%%%%%%%%%%%%%%
%% References
%%%%%%%%%%%%%%%%%%%%%%%%%%%%%%%%%%%%%%%%%%%%%%%%%%%%%%%%%%%%%%%%%%%%%%%%%%%%%%%%

{\small
\bibliographystyle{abbrvnat}
\bibliography{main}

@inproceedings{yu2025objectmover,
  title={Objectmover: Generative object movement with video prior},
  author={Yu, Xin and Wang, Tianyu and Kim, Soo Ye and Guerrero, Paul and Chen, Xi and Liu, Qing and Lin, Zhe and Qi, Xiaojuan},
  booktitle={Proceedings of the IEEE/CVF Conference on Computer Vision and Pattern Recognition},
  pages={17682--17691},
  year={2025}
}

@article{wang2025frame,
  title={Frame in-n-out: Unbounded controllable image-to-video generation},
  author={Wang, Boyang and Chen, Xuweiyi and Gadelha, Matheus and Cheng, Zezhou},
  journal={arXiv preprint arXiv:2505.21491},
  year={2025}
}

@article{miao2025rose,
  title={Rose: Remove objects with side effects in videos},
  author={Miao, Chenxuan and Feng, Yutong and Zeng, Jianshu and Gao, Zixiang and Liu, Hantang and Yan, Yunfeng and Qi, Donglian and Chen, Xi and Wang, Bin and Zhao, Hengshuang},
  journal={arXiv preprint arXiv:2508.18633},
  year={2025}
}

@article{gao2026pisco,
  title={Pisco: Precise video instance insertion with sparse control},
  author={Gao, Xiangbo and Li, Renjie and Chen, Xinghao and Wu, Yuheng and Feng, Suofei and Yin, Qing and Tu, Zhengzhong},
  journal={arXiv preprint arXiv:2602.08277},
  year={2026}
}

@inproceedings{jiang2025vace,
  title={Vace: All-in-one video creation and editing},
  author={Jiang, Zeyinzi and Han, Zhen and Mao, Chaojie and Zhang, Jingfeng and Pan, Yulin and Liu, Yu},
  booktitle={Proceedings of the IEEE/CVF International Conference on Computer Vision},
  pages={17191--17202},
  year={2025}
}

@article{wei2025univideo,
  title={Univideo: Unified understanding, generation, and editing for videos},
  author={Wei, Cong and Liu, Quande and Ye, Zixuan and Wang, Qiulin and Wang, Xintao and Wan, Pengfei and Gai, Kun and Chen, Wenhu},
  journal={arXiv preprint arXiv:2510.08377},
  year={2025}
}

@inproceedings{bian2025videopainter,
  title={Videopainter: Any-length video inpainting and editing with plug-and-play context control},
  author={Bian, Yuxuan and Zhang, Zhaoyang and Ju, Xuan and Cao, Mingdeng and Xie, Liangbin and Shan, Ying and Xu, Qiang},
  booktitle={Proceedings of the Special Interest Group on Computer Graphics and Interactive Techniques Conference Conference Papers},
  pages={1--12},
  year={2025}
}

@inproceedings{zi2025cococo,
  title={Cococo: Improving text-guided video inpainting for better consistency, controllability and compatibility},
  author={Zi, Bojia and Zhao, Shihao and Qi, Xianbiao and Wang, Jianan and Shi, Yukai and Chen, Qianyu and Liang, Bin and Xiao, Rong and Wong, Kam-Fai and Zhang, Lei},
  booktitle={Proceedings of the AAAI Conference on Artificial Intelligence},
  volume={39},
  number={10},
  pages={11067--11076},
  year={2025}
}

@article{fu2026effecterase,
  title={EffectErase: Joint Video Object Removal and Insertion for High-Quality Effect Erasing},
  author={Fu, Yang and Zheng, Yike and Dai, Ziyun and Ding, Henghui},
  journal={arXiv preprint arXiv:2603.19224},
  year={2026}
}

@article{jin2025insertanywhere,
  title={InsertAnywhere: Bridging 4D Scene Geometry and Diffusion Models for Realistic Video Object Insertion},
  author={Jin, Hoiyeong and Jang, Hyojin and Kim, Jeongho and Hyung, Junha and Kim, Kinam and Kim, Dongjin and Choi, Huijin and Kim, Hyeonji and Choo, Jaegul},
  journal={arXiv preprint arXiv:2512.17504},
  year={2025}
}

@article{motamed2026void,
  title={VOID: Video Object and Interaction Deletion},
  author={Motamed, Saman and Harvey, William and Klein, Benjamin and Van Gool, Luc and Yuan, Zhuoning and Cheng, Ta-Ying},
  journal={arXiv preprint arXiv:2604.02296},
  year={2026}
}

@article{fang2025v,
  title={V-RGBX: Video Editing with Accurate Controls over Intrinsic Properties},
  author={Fang, Ye and Wu, Tong and Deschaintre, Valentin and Ceylan, Duygu and Georgiev, Iliyan and Huang, Chun-Hao Paul and Hu, Yiwei and Chen, Xuelin and Wang, Tuanfeng Yang},
  journal={arXiv preprint arXiv:2512.11799},
  year={2025}
}

@inproceedings{zeng2024rgb,
author = {Zeng, Zheng and Deschaintre, Valentin and Georgiev, Iliyan and Hold-Geoffroy, Yannick and Hu, Yiwei and Luan, Fujun and Yan, Ling-Qi and Ha\v{s}an, Milo\v{s}},
title = {RGB<->X: Image decomposition and synthesis using material- and lighting-aware diffusion models},
year = {2024},
isbn = {9798400705250},
publisher = {Association for Computing Machinery},
address = {New York, NY, USA},
url = {https://doi.org/10.1145/3641519.3657445},
doi = {10.1145/3641519.3657445},
booktitle = {ACM SIGGRAPH 2024 Conference Papers},
articleno = {75},
numpages = {11},
location = {Denver, CO, USA},
series = {SIGGRAPH '24}
}

@inproceedings{liang2025diffusion,
  title={Diffusion renderer: Neural inverse and forward rendering with video diffusion models},
  author={Liang, Ruofan and Gojcic, Zan and Ling, Huan and Munkberg, Jacob and Hasselgren, Jon and Lin, Chih-Hao and Gao, Jun and Keller, Alexander and Vijaykumar, Nandita and Fidler, Sanja and others},
  booktitle={Proceedings of the Computer Vision and Pattern Recognition Conference},
  pages={26069--26080},
  year={2025}
}

@article{he2025unirelight,
  title={Unirelight: Learning joint decomposition and synthesis for video relighting},
  author={He, Kai and Liang, Ruofan and Munkberg, Jacob and Hasselgren, Jon and Vijaykumar, Nandita and Keller, Alexander and Fidler, Sanja and Gilitschenski, Igor and Gojcic, Zan and Wang, Zian},
  journal={arXiv preprint arXiv:2506.15673},
  year={2025}
}

@article{wan2025wan,
  title={Wan: Open and advanced large-scale video generative models},
  author={Wan, Team and Wang, Ang and Ai, Baole and Wen, Bin and Mao, Chaojie and Xie, Chen-Wei and Chen, Di and Yu, Feiwu and Zhao, Haiming and Yang, Jianxiao and others},
  journal={arXiv preprint arXiv:2503.20314},
  year={2025}
}

@article{bai2023qwen,
  title={Qwen technical report},
  author={Bai, Jinze and Bai, Shuai and Chu, Yunfei and Cui, Zeyu and Dang, Kai and Deng, Xiaodong and Fan, Yang and Ge, Wenbin and Han, Yu and Huang, Fei and others},
  journal={arXiv preprint arXiv:2309.16609},
  year={2023}
}

@inproceedings{deitke2023objaverse,
  title={Objaverse: A universe of annotated 3d objects},
  author={Deitke, Matt and Schwenk, Dustin and Salvador, Jordi and Weihs, Luca and Michel, Oscar and VanderBilt, Eli and Schmidt, Ludwig and Ehsani, Kiana and Kembhavi, Aniruddha and Farhadi, Ali},
  booktitle={Proceedings of the IEEE/CVF conference on computer vision and pattern recognition},
  pages={13142--13153},
  year={2023}
}

@article{lipman2022flow,
  title={Flow matching for generative modeling},
  author={Lipman, Yaron and Chen, Ricky TQ and Ben-Hamu, Heli and Nickel, Maximilian and Le, Matt},
  journal={arXiv preprint arXiv:2210.02747},
  year={2022}
}

@inproceedings{rombach2022high,
  title={High-resolution image synthesis with latent diffusion models},
  author={Rombach, Robin and Blattmann, Andreas and Lorenz, Dominik and Esser, Patrick and Ommer, Bj{\"o}rn},
  booktitle={Proceedings of the IEEE/CVF conference on computer vision and pattern recognition},
  pages={10684--10695},
  year={2022}
}

@inproceedings{peebles2023scalable,
  title={Scalable diffusion models with transformers},
  author={Peebles, William and Xie, Saining},
  booktitle={Proceedings of the IEEE/CVF international conference on computer vision},
  pages={4195--4205},
  year={2023}
}

@inproceedings{radford2021learning,
  title={Learning transferable visual models from natural language supervision},
  author={Radford, Alec and Kim, Jong Wook and Hallacy, Chris and Ramesh, Aditya and Goh, Gabriel and Agarwal, Sandhini and Sastry, Girish and Askell, Amanda and Mishkin, Pamela and Clark, Jack and others},
  booktitle={International conference on machine learning},
  pages={8748--8763},
  year={2021},
  organization={PmLR}
}

@article{hacohen2026ltx,
  title={LTX-2: Efficient Joint Audio-Visual Foundation Model},
  author={HaCohen, Yoav and Brazowski, Benny and Chiprut, Nisan and Bitterman, Yaki and Kvochko, Andrew and Berkowitz, Avishai and Shalem, Daniel and Lifschitz, Daphna and Moshe, Dudu and Porat, Eitan and others},
  journal={arXiv preprint arXiv:2601.03233},
  year={2026}
}

@article{kong2024hunyuanvideo,
  title={Hunyuanvideo: A systematic framework for large video generative models},
  author={Kong, Weijie and Tian, Qi and Zhang, Zijian and Min, Rox and Dai, Zuozhuo and Zhou, Jin and Xiong, Jiangfeng and Li, Xin and Wu, Bo and Zhang, Jianwei and others},
  journal={arXiv preprint arXiv:2412.03603},
  year={2024}
}

@article{yang2024cogvideox,
  title={CogVideoX: Text-to-Video Diffusion Models with An Expert Transformer},
  author={Yang, Zhuoyi and Teng, Jiayan and Zheng, Wendi and Ding, Ming and Huang, Shiyu and Xu, Jiazheng and Yang, Yuanming and Hong, Wenyi and Zhang, Xiaohan and Feng, Guanyu and others},
  journal={arXiv preprint arXiv:2408.06072},
  year={2024}
}

@article{hong2022cogvideo,
  title={Cogvideo: Large-scale pretraining for text-to-video generation via transformers},
  author={Hong, Wenyi and Ding, Ming and Zheng, Wendi and Liu, Xinghan and Tang, Jie},
  journal={arXiv preprint arXiv:2205.15868},
  year={2022}
}

@article{wiedemer2025video,
  title={Video models are zero-shot learners and reasoners},
  author={Wiedemer, Thadd{\"a}us and Li, Yuxuan and Vicol, Paul and Gu, Shixiang Shane and Matarese, Nick and Swersky, Kevin and Kim, Been and Jaini, Priyank and Geirhos, Robert},
  journal={arXiv preprint arXiv:2509.20328},
  year={2025}
}

@inproceedings{abdal2025dynamic,
  title={Dynamic concepts personalization from single videos},
  author={Abdal, Rameen and Patashnik, Or and Skorokhodov, Ivan and Menapace, Willi and Siarohin, Aliaksandr and Tulyakov, Sergey and Cohen-Or, Daniel and Aberman, Kfir},
  booktitle={Proceedings of the Special Interest Group on Computer Graphics and Interactive Techniques Conference Conference Papers},
  pages={1--9},
  year={2025}
}

@article{garibi2025tokenverse,
  title={Tokenverse: Versatile multi-concept personalization in token modulation space},
  author={Garibi, Daniel and Yadin, Shahar and Paiss, Roni and Tov, Omer and Zada, Shiran and Ephrat, Ariel and Michaeli, Tomer and Mosseri, Inbar and Dekel, Tali},
  journal={ACM Transactions On Graphics (TOG)},
  volume={44},
  number={4},
  pages={1--11},
  year={2025},
  publisher={ACM New York, NY, USA}
}

@inproceedings{gu2025diffusion,
  title={Diffusion as shader: 3d-aware video diffusion for versatile video generation control},
  author={Gu, Zekai and Yan, Rui and Lu, Jiahao and Li, Peng and Dou, Zhiyang and Si, Chenyang and Dong, Zhen and Liu, Qifeng and Lin, Cheng and Liu, Ziwei and others},
  booktitle={Proceedings of the Special Interest Group on Computer Graphics and Interactive Techniques Conference Conference Papers},
  pages={1--12},
  year={2025}
}

@article{lyu2025intrinsicedit,
  title={IntrinsicEdit: Precise generative image manipulation in intrinsic space},
  author={Lyu, Linjie and Deschaintre, Valentin and Hold-Geoffroy, Yannick and Ha{\v{s}}an, Milo{\v{s}} and Yoon, Jae Shin and Leimk{\"u}hler, Thomas and Theobalt, Christian and Georgiev, Iliyan},
  journal={ACM Transactions on Graphics (TOG)},
  volume={44},
  number={4},
  pages={1--13},
  year={2025},
  publisher={ACM New York, NY, USA}
}

@article{shin2025motionstream,
  title={Motionstream: Real-time video generation with interactive motion controls},
  author={Shin, Joonghyuk and Li, Zhengqi and Zhang, Richard and Zhu, Jun-Yan and Park, Jaesik and Shechtman, Eli and Huang, Xun},
  journal={arXiv preprint arXiv:2511.01266},
  year={2025}
}

@article{lee2025generative,
  title={Generative Video Motion Editing with 3D Point Tracks},
  author={Lee, Yao-Chih and Zhang, Zhoutong and Huang, Jiahui and Wang, Jui-Hsien and Lee, Joon-Young and Huang, Jia-Bin and Shechtman, Eli and Li, Zhengqi},
  journal={arXiv preprint arXiv:2512.02015},
  year={2025}
}

@article{dhariwal2021diffusion,
  title={Diffusion models beat gans on image synthesis},
  author={Dhariwal, Prafulla and Nichol, Alexander},
  journal={Advances in neural information processing systems},
  volume={34},
  pages={8780--8794},
  year={2021}
}

@article{ho2020denoising,
  title={Denoising diffusion probabilistic models},
  author={Ho, Jonathan and Jain, Ajay and Abbeel, Pieter},
  journal={Advances in neural information processing systems},
  volume={33},
  pages={6840--6851},
  year={2020}
}

@article{van2017neural,
  title={Neural discrete representation learning},
  author={Van Den Oord, Aaron and Vinyals, Oriol and others},
  journal={Advances in neural information processing systems},
  volume={30},
  year={2017}
}

@article{kingma2013auto,
  title={Auto-encoding variational bayes},
  author={Kingma, Diederik P and Welling, Max},
  journal={arXiv preprint arXiv:1312.6114},
  year={2013}
}

@misc{photoshop,
  author = {{Adobe Inc.}},
  title = {Adobe Photoshop},
  year = {2026},
  howpublished = {\url{https://www.adobe.com/products/photoshop.html}}
}

@article{ye2025unic,
  title={Unic: Unified in-context video editing},
  author={Ye, Zixuan and He, Xuanhua and Liu, Quande and Wang, Qiulin and Wang, Xintao and Wan, Pengfei and Zhang, Di and Gai, Kun and Chen, Qifeng and Luo, Wenhan},
  journal={arXiv preprint arXiv:2506.04216},
  year={2025}
}

@article{bai2024anything,
  title={Anything in any scene: Photorealistic video object insertion},
  author={Bai, Chen and Shao, Zeman and Zhang, Guoxiang and Liang, Di and Yang, Jie and Zhang, Zhuorui and Guo, Yujian and Zhong, Chengzhang and Qiu, Yiqiao and Wang, Zhendong and others},
  journal={arXiv preprint arXiv:2401.17509},
  year={2024}
}

@article{saini2024invi,
  title={Invi: Object insertion in videos using off-the-shelf diffusion models},
  author={Saini, Nirat and Bodla, Navaneeth and Shrivastava, Ashish and Ravichandran, Avinash and Zhang, Xiao and Shrivastava, Abhinav and Singh, Bharat},
  journal={arXiv preprint arXiv:2407.10958},
  year={2024}
}

@article{zhao2025dreaminsert,
  title={Dreaminsert: Zero-shot image-to-video object insertion from a single image},
  author={Zhao, Qi and Ma, Zhan and Zhou, Pan},
  journal={arXiv preprint arXiv:2503.10342},
  year={2025}
}

@article{batifol2025flux,
  title={Flux. 1 kontext: Flow matching for in-context image generation and editing in latent space},
  author={Batifol, Stephen and Blattmann, Andreas and Boesel, Frederic and Consul, Saksham and Diagne, Cyril and Dockhorn, Tim and English, Jack and English, Zion and Esser, Patrick and Kulal, Sumith and others},
  journal={arXiv e-prints},
  pages={arXiv--2506},
  year={2025}
}

@article{zhang2026diffusionharmonizer,
  title={DiffusionHarmonizer: Bridging Neural Reconstruction and Photorealistic Simulation with Online Diffusion Enhancer},
  author={Zhang, Yuxuan and T{\'o}thov{\'a}, Katar{\'\i}na and Wang, Zian and Yin, Kangxue and Turki, Haithem and de Lutio, Riccardo and Chang, Yen-Yu and Litany, Or and Fidler, Sanja and Gojcic, Zan},
  journal={arXiv preprint arXiv:2602.24096},
  year={2026}
}

@inproceedings{zhou2023propainter,
  title={Propainter: Improving propagation and transformer for video inpainting},
  author={Zhou, Shangchen and Li, Chongyi and Chan, Kelvin CK and Loy, Chen Change},
  booktitle={Proceedings of the IEEE/CVF international conference on computer vision},
  pages={10477--10486},
  year={2023}
}

@article{ku2024anyv2v,
  title={Anyv2v: A tuning-free framework for any video-to-video editing tasks},
  author={Ku, Max and Wei, Cong and Ren, Weiming and Yang, Harry and Chen, Wenhu},
  journal={arXiv preprint arXiv:2403.14468},
  year={2024}
}

@inproceedings{yu2025trajectorycrafter,
  title={Trajectorycrafter: Redirecting camera trajectory for monocular videos via diffusion models},
  author={Yu, Mark and Hu, Wenbo and Xing, Jinbo and Shan, Ying},
  booktitle={Proceedings of the IEEE/CVF international conference on computer vision},
  pages={100--111},
  year={2025}
}

@article{he2024cameractrl,
  title={Cameractrl: Enabling camera control for text-to-video generation},
  author={He, Hao and Xu, Yinghao and Guo, Yuwei and Wetzstein, Gordon and Dai, Bo and Li, Hongsheng and Yang, Ceyuan},
  journal={arXiv preprint arXiv:2404.02101},
  year={2024}
}

@inproceedings{zhang2023adding,
  title={Adding conditional control to text-to-image diffusion models},
  author={Zhang, Lvmin and Rao, Anyi and Agrawala, Maneesh},
  booktitle={Proceedings of the IEEE/CVF international conference on computer vision},
  pages={3836--3847},
  year={2023}
}

@article{li2025diffueraser,
  title={Diffueraser: A diffusion model for video inpainting},
  author={Li, Xiaowen and Xue, Haolan and Ren, Peiran and Bo, Liefeng},
  journal={arXiv preprint arXiv:2501.10018},
  year={2025}
}

@article{loshchilov2017decoupled,
  title={Decoupled weight decay regularization},
  author={Loshchilov, Ilya and Hutter, Frank},
  journal={arXiv preprint arXiv:1711.05101},
  year={2017}
}

@InProceedings{lin2025objaverseplusplus,
    author    = {Lin, Chendi and Liu, Heshan and Lin, Qunshu and Bright, Zachary and Tang, Shitao and He, Yihui and Liu, Minghao and Zhu, Ling and Le, Cindy},
    title     = {{Objaverse++: Curated 3D Object Dataset with Quality Annotations}},
    booktitle = {Proceedings of the IEEE/CVF International Conference on Computer Vision (ICCV) Workshops},
    year      = {2025},
    pages     = {6813-6822}
}

@article{ma2025materialpicker,
   title={MaterialPicker: Multi-Modal DiT-Based Material Generation},
   volume={44},
   ISSN={1557-7368},
   url={http://dx.doi.org/10.1145/3731199},
   DOI={10.1145/3731199},
   number={4},
   journal={ACM Transactions on Graphics},
   publisher={Association for Computing Machinery (ACM)},
   author={Ma, Xiaohe and Deschaintre, Valentin and Hašan, Miloš and Luan, Fujun and Zhou, Kun and Wu, Hongzhi and Hu, Yiwei},
   year={2025},
   month=jul, pages={1–12}
}

@article{wang2004image,
  title={Image quality assessment: from error visibility to structural similarity},
  author={Wang, Zhou and Bovik, Alan C and Sheikh, Hamid R and Simoncelli, Eero P},
  journal={IEEE transactions on image processing},
  volume={13},
  number={4},
  pages={600--612},
  year={2004},
  publisher={IEEE}
}

@inproceedings{zhang2018unreasonable,
  title={The unreasonable effectiveness of deep features as a perceptual metric},
  author={Zhang, Richard and Isola, Phillip and Efros, Alexei A and Shechtman, Eli and Wang, Oliver},
  booktitle={Proceedings of the IEEE conference on computer vision and pattern recognition},
  pages={586--595},
  year={2018}
}

@InProceedings{huang2023vbench,
    title={{VBench}: Comprehensive Benchmark Suite for Video Generative Models},
    author={Huang, Ziqi and He, Yinan and Yu, Jiashuo and Zhang, Fan and Si, Chenyang and Jiang, Yuming and Zhang, Yuanhan and Wu, Tianxing and Jin, Qingyang and Chanpaisit, Nattapol and Wang, Yaohui and Chen, Xinyuan and Wang, Limin and Lin, Dahua and Qiao, Yu and Liu, Ziwei},
    booktitle={Proceedings of the IEEE/CVF Conference on Computer Vision and Pattern Recognition},
    year={2024}
}

@inproceedings{sohl2015deep,
  title={Deep unsupervised learning using nonequilibrium thermodynamics},
  author={Sohl-Dickstein, Jascha and Weiss, Eric and Maheswaranathan, Niru and Ganguli, Surya},
  booktitle={International conference on machine learning},
  pages={2256--2265},
  year={2015},
  organization={pmlr}
}
}

%%%%%%%%%%%%%%%%%%%%%%%%%%%%%%%%%%%%%%%%%%%%%%%%%%%%%%%%%%%%%%%%%%%%%%%%%%%%%%%%
%% Supplemental / extra figures (placed after references, as in original)
%%%%%%%%%%%%%%%%%%%%%%%%%%%%%%%%%%%%%%%%%%%%%%%%%%%%%%%%%%%%%%%%%%%%%%%%%%%%%%%%

% \begin{figure*}[p]
%   \centering
%   \begin{minipage}[t]{0.48\textwidth}
%     \centering
%     \includegraphics[width=\linewidth]{fig/Limitation_c.pdf}
%     \vspace{-0.2in}
%     \caption{Limitation of our method. As shown in the figure, when side effects of edited objects are mistakenly counted as texture (highlighted by the red arrow) in the albedo maps, our method fails to remove the corresponding side effects in the edited videos.}
%     \label{fig:limitation}
%   \end{minipage}
%   \hfill
%   \begin{minipage}[t]{0.48\textwidth}
%     \centering
%     \includegraphics[width=\linewidth]{fig/Ablation_c.pdf}
%     \vspace{-0.2in}
%     \caption{Left: We demonstrate the effect of inserting objects that should plausibly be emissive; the model correctly infers this. Right: using a smaller base model (Wan 1.3B) on the insertion task leads to noticeably lower quality.}
%     \label{fig:ablation}
%   \end{minipage}
% \end{figure*}

\end{document}